\documentclass[11pt,a4paper]{article}

\usepackage{graphicx} 
\usepackage{jheppub} 

\def\Pom{{\bf I\!P}}
\def\Reg{{\bf I\!R}}
\newcommand{\p}{\partial}

\newcommand{\twosidep}[1]{\stackrel{\leftrightarrow}{\p^{#1}}}

\keywords{Phenomenological Models, QCD Phenomenology}

\title{
$W^+ W^-$ pair production in proton-proton collisions: small missing terms}

\author[a]{Marta {\L}uszczak}
\author[a,b]{Antoni Szczurek}
\author[c]{Christophe Royon}

\affiliation[a]{University of Rzesz\'ow, PL-35-959 Rzesz\'ow, Poland}
\affiliation[b]{Institute of Nuclear Physics PAN, PL-31-342 Krak\'ow, Poland}
\affiliation[c]{IRFU/Service de Physique des Particules, CEA/Saclay, France}

\emailAdd{luszczak@univ.rzeszow.pl}
\emailAdd{antoni.szczurek@ifj.edu.pl}
\emailAdd{christophe.royon@cea.fr}

\abstract{
$W^+ W^-$ production is one of the golden channels for testing the
Standard Model as well for searches beyond the Standard Model.
We discuss many new subleading processes for inclusive production of 
$W^+ W^-$ pairs generally not included in the litterature so far. 
We focus mainly on photon-photon induced processes.
We include elastic-elastic, elastic-inelastic, inelastic-elastic 
and inelastic-inelastic contributions. 
We also calculate the contributions with resolved photons including 
the partonic substructure of the virtual photon. 
Predictions for the total cross section and differential
distributions in $W$- boson rapidity and transverse momentum as well
as $WW$ invariant mass are presented. The $\gamma \gamma$ components
only constitute about 1-2 \% of the inclusive $W^+ W^-$ cross section
but increases up to 
about 10 \% at large $W^{\pm}$ transverse momenta, and are even
comparable to the dominant $q \bar q$ component at large $M_{WW}$,
i.e. are much larger than the $g g \to W^+ W^-$ one.
}

\begin{document}

\maketitle

\flushbottom

\section{Introduction}
\label{intro}

The  $p p \to W^+ W^- X$ process is quite fundamental in particle physics.
It constitutes an important, irreducible
background to the observation of the Higgs boson in the $W^+ W^-$
channel and furthermore,
can be used to test Standard Model gauge boson couplings
and study them in models beyond the Standard Model.

Especially, the exclusive $W^+ W^-$ process is interesting 
by itself since it can be used to test the Standard Model and many  
beyond Standard Model theories.
The photon-photon contribution was recently considered in the litterature 
\cite{royon,piotrzkowski} and dominates at $W$ pair masses.
The exclusive QCD diffractive mechanism of central exclusive production
of $W^+W^-$ pairs 
(in which diagrams with intermediate virtual Higgs boson as well as quark box
diagrams are included) was discussed in Ref.~\cite{LS2012} and turned 
out to  be small at high masses but dominating at low masses.
The $W^+W^-$ pair production signal 
is particularly sensitive to new Physics contributions in 
the $\gamma \gamma \to W^+ W^-$ subprocess \cite{royon,piotrzkowski}. 
Similar analysis has been considered recently
for $\gamma \gamma \to Z Z$ and $\gamma \gamma \to \gamma \gamma$ 
\cite{Gupta:2011be, royon}. 
Corresponding measurements would be possible to perform at ATLAS or CMS
provided the very forward proton detectors are installed
\cite{forward_protons}. 

In the present paper we concentrate on the inclusive and exclusive productions 
of $W^+ W^-$ pairs.
The inclusive production of $W^+ W^-$ has been measured recently by 
the CMS and ATLAS collaborations \cite{CMS2011, ATLAS2012}.
The total measured cross section by the CMS collaboration is  
41.1 $\pm$ 15.3 (stat) $\pm$ 5.8 (syst) $\pm$ 4.5 (lumi) pb, 
the total measured cross section using the ATLAS detector with slightly
better statistics is
54.4 $\pm$  4.0 (stat.) $\pm$  3.9 (syst.) $\pm$  2.0 (lumi.) pb. 
The  ATLAS and CMS results are somewhat larger than the Standard Model
predictions of 44.4  $\pm$  2.8 pb \cite{ATLAS2012}.
The Standard Model predictions do not include several potentially
important subleading processes, and this is the aim of this paper to compute the
usually missing processes in the litterature leading to pairs of $W$s in the
final state.

In this paper, we  review several  processes which
are usually ignored in the litterature and discuss if they could explain
the slight discrepancy between the measurements and the usual SM predictions.
Some of the not included processes were already
discussed previously. One of such examples is the double parton
scattering (DPS) that
was discussed e.g. in Ref.\cite{KS2000,Kulesza2010,GKKS2011}. 
The $W^+ W^-$ final states constitutes a background to the Higgs boson 
production.

In section II of the paper, we will review the different processes
leading to two $W$s in the final state and compute the production cross
sections. We will discuss in turn the exclusive production of $W$ pairs via
photon and gluon exchanges, the inclusive one originating by quark and gluon
exchanges, the inelastic-inelastic and elastic-inelastic productions, 
as well as the diffractive contributions.
We thus include for the first time processes with resolved 
photons as well as single-diffractive production of $W^+ W^-$ pairs.
In section
III, we discuss the results and study how to measure each component
individually.

\section{Processes leading to $W$ pair production}
\label{leading}

In this section, we review all processes leading to pairs of $W$s in the final state.

\subsection{$\gamma \gamma \to W^+ W^-$ reaction}

In this section, we discuss a  basic ingredient to produce $W$ pairs exclusively
via photon exchanges. In the next section, we will convolute this elementary
cross section with the photon flux originating from the protons.

Let us start from a reminder about the $\gamma \gamma \to W^+ W^-$
coupling within the Standard Model.
The three-boson $WW \gamma$ and four-boson $WW \gamma\gamma$ couplings,
which contribute to the $\gamma \gamma \to W^+ W^-$ process in
the leading order read
\begin{eqnarray}
{\cal L}_{WW\gamma} & = &
-ie( A_\mu  W^-_\nu \twosidep{\mu} W^{+\nu}
+   W_\mu^- W^+_\nu \twosidep{\mu} A^\nu
+    W^+_\mu  A_\nu \twosidep{\mu} W^{-\nu}) \, ,
\nonumber \\ 
{\cal L}_{WW\gamma\gamma} & = &
-e^2\left(  W^{-}_{\mu} W^{+\mu}A_{\nu} A^{\nu} 
     - W^{+}_{\mu}A^{\mu} W^{-}_{\nu} A^\nu \right) \, ,
\label{eq:anom:lagrww2}
\end{eqnarray}
where the asymmetric derivative has the form
$X\twosidep{\mu}Y=X\p^{\mu}Y-Y\p^{\mu}X$.

\begin{figure*}
\begin{center}
\includegraphics[width=40mm,height=40mm]{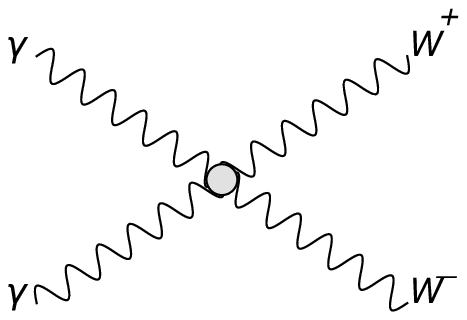}
\includegraphics[width=40mm,height=40mm]{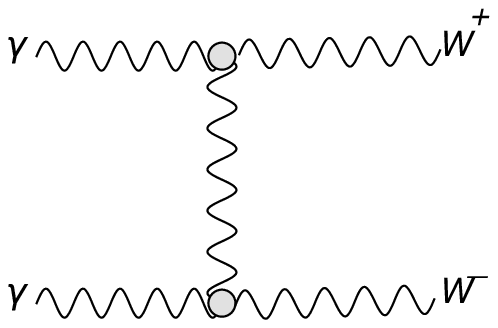}
\includegraphics[width=40mm,height=40mm]{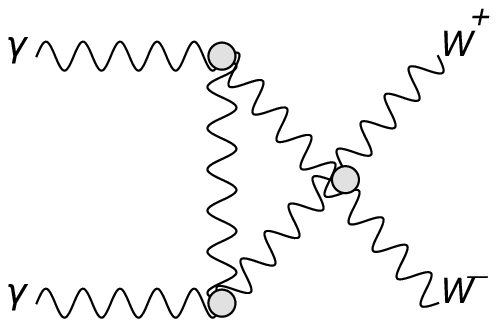} 
\end{center}
\caption{The leading order $\gamma \gamma \to W^+ W^-$ subprocesses.
}
\label{fig:LO_subprocesses}
\end{figure*}

The relevant leading-order subprocess diagrams are shown 
in Fig.~\ref{fig:LO_subprocesses}.

Within the Standard Model, the elementary tree-level cross
section for the $\gamma \gamma \to W^+ W^-$ subprocess can be written 
in the very compact form in terms of the Mandelstam variables 
(see e.g. Ref.~\cite{DDS95})

\begin{equation}
\frac{d\hat{\sigma}}{d \Omega} = \frac{3 \alpha^2 \beta}{2\hat{s}} \left(
1 - \frac{2 \hat{s} (2\hat{s}+3m_W^2)}{3 (m_W^2 - \hat{t}) (m_W^2 -
\hat{u})} + \frac{2 \hat{s}^2(\hat{s}^2+ 3m_W^4)}{3 (m_W^2 -
\hat{t})^2(m_W^2 - \hat{u})^2} \right) \, ,
\label{gamgam_WW}
\end{equation}
where $\beta=\sqrt{1-4m_W^2/\hat{s}}$ is the velocity of the $W$
bosons in their center-of-mass frame and the electromagnetic
fine-structure constant $\alpha=e^{2}/(4\pi) \simeq 1/137$. 
The total elementary cross section can be obtained
by integration of the differential cross section given above.

\subsection{Exclusive $p p \to p p W^+ W^-$ reaction}

In this section, we give the exclusive $WW$ production cross section via photon
exchanges.

\begin{figure*}
\begin{center}
\includegraphics[width=5cm,height=5cm]{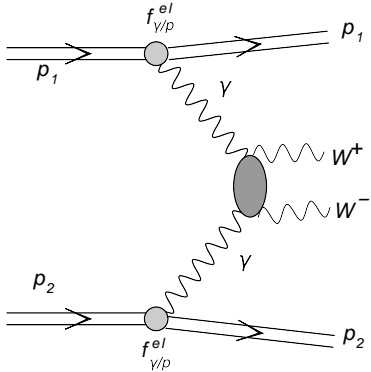}
\caption{The general diagram for the $pp \to pp W^+ W^-$ reaction
via $\gamma \gamma \to W^+ W^-$ subprocess.
}
\label{fig:m1}
\end{center}
\end{figure*}

The $p p \to p p W^+ W^-$ reaction is particularly interesting in the
context of $\gamma \gamma W W$ coupling \cite{royon,piotrzkowski}.
The general diagram for the exclusive reaction is shown in
Fig.~\ref{fig:m1}.

In the Weizs\"acker-Williams approximation, which was used so far in the
litterature,
the total cross section for the $pp \to pp (\gamma \gamma) \to W^+ W^-$
can be written as in the parton model:
\begin{equation}
\sigma = \int d x_1 d x_2 \, f_1(x_1) \, f_2(x_2) \,
\hat{\sigma}_{\gamma \gamma \to W^+ W^-}(\hat s) \, .
\label{EPA}
\end{equation}
We take the Weizs\"acker-Williams equivalent photon fluxes $f_1$ and $f_2$ in
protons from Ref.~\cite{DZ}.

To calculate differential distributions the following parton formula
can be conveniently used
\begin{equation}
\frac{d\sigma}{d y_+ d y_- d^2 p_{W\perp}} = \frac{1}{16 \pi^2 {\hat s}^2}
\, x_1 f_1(x_1) \, x_2 f_2(x_2) \,
\overline{ | {\cal M}_{\gamma \gamma \to W^+ W^-}(\hat s, \hat t, \hat u)
  |^2} \, .
\label{EPA_differential}
\end{equation}
We shall not discuss here any approach beyond the Standard Model.
A potentially interesting beyond Standard Model production of 
$W^+ W^-$ pairs due to the existence of extra-dimensions has been discussed 
previously e.g. in
Refs.~\cite{royon,piotrzkowski}.

The exclusive cross section could be also calculated more precisely
in four-body calculations using the corresponding $2 \to 4$ matrix element.
If exact exclusivity of the reaction is required then absorption effect
related to soft proton-proton interactions has to be included. Such
effects can be included consistently only in the 
four-body calculations but can be included only approximately
in the Weizsacker-Williams approximation by multiplying the Born cross
section by a given factor. From our experience in other 
$\gamma \gamma$ processes where four-body calculations were done
this factor is about 0.8-0.9, that is of the order of 1, 
depending on the mass of the produced
system. In the following, we assume this factor to be 1 and all cross sections
should be multiplied by this factor once known.

\subsection{Inclusive production of $W^+W^-$ pairs}
\label{sec:inclusive}

In this section, we discuss the inclusive production of $W$ pairs. We first
start by the usual production considered in the litterature via gluon and quark
exchanges, We then describe new processes, starting by the $W$ pair production
via photon exchanges where we distinguish the elastic-elastic, 
elastic-inelastic and inelastic-inelastic contributions. 
We finish the section by studying the diffractive production of $W$
pairs. 
All these contributions are ``inclusive" in
the sense that additional particles are produced together with the $W$ pair.

The dominant contribution of $W^+W^-$ pair production is initiated by
quark-antiquark annihilation \cite{DDS95}.
The gluon-gluon contribution to the inclusive cross section 
was calculated first in Ref.~\cite{gg_WW}.

Therefore, for reference, we also consider the quark-antiquark and 
gluon-gluon components to the inclusive cross section. 
We briefly remind the basics for
these well known dominant contributions.

\subsubsection{$q \bar q \to W^+ W^-$ mechanism}

\begin{figure*}
\begin{center}
\includegraphics[width=5cm,height=5cm]{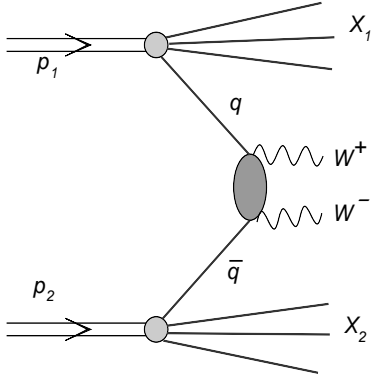}
\caption{A generic diagram representing mechanisms 
for the production of $W^+W^-$ pairs in the
$q \bar q \to W^+ W^-$ subprocess.
}
\label{fig:qqbar_WW}
\end{center}
\end{figure*}

The generic diagram for the $q \bar q$ initiated processes
is shown in Fig.\ref{fig:qqbar_WW}. This contains $t$- and $u$-channel quark
exchanges as well as $s$-channel photon and $Z$-boson exchanges
\cite{DDS95}.

Therefore this process is also of interest as a probe of 
the gauge structure of the electroweak interactions.
Relevant leading-order matrix element, averaged over quark colors 
and over initial spin polarizations and summed over final spin 
polarization, can be found e.g. in Ref.~\cite{Eichten}.

The corresponding differential cross section in leading-order
approximation can be calculated as:
\begin{equation}
\frac{d\sigma}{d y_+ d y_- d^2 p_{W\perp}} = \frac{1}{16 \pi^2 {\hat s}^2}
\,\sum_{f}[ x_1 q_f(x_1,\mu^2) \, x_2 {\bar q_f}(x_2,\mu^2) +
    x_1 {\bar q_f}(x_1,\mu^2) \, x_2 q_f(x_2,\mu^2) ] \; \nonumber \\
 \overline{ | {\cal M}_{q {\bar q} \to W^+ W^-}(\hat s, \hat t, \hat u)
  |^2} \, .
\label{qqbar_annihilation}
\end{equation}
Above $q_f$ and $\bar q_f$ are quark and antiquark distributions of a given
flavor in the proton, $x_1$ and $x_2$ are corresponding longitudinal momentum
fractions carried by the quark or antiquark and $\mu^2$ is a QCD scale taken here
to be $\mu^2 = m_t^2$, where $m_t$ is the W-boson transverse mass.

\subsubsection{$g g \to W^+ W^-$ mechanism}

\begin{figure*}
\begin{center}
\includegraphics[width=5cm,height=5cm]{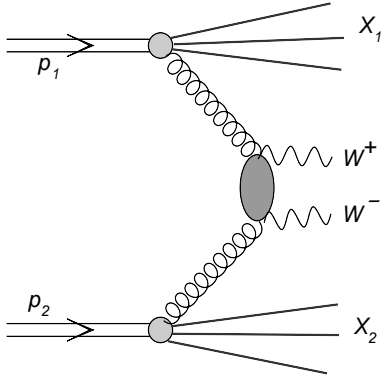}
\caption{A generic diagram representing mechanisms 
for the production of $W^+W^-$ pairs in the
$gg \to W^+ W^-$ subprocess.
}
\label{fig:gg_WW}
\end{center}
\end{figure*}

The generic diagram for the $g g$ initiated processes 
is shown in Fig.\ref{fig:gg_WW}. This contains both quark box diagrams
and heavy-quark triangle with s-channel Higgs boson in the
intermediate stage. More details of the relevant calculation can be
found e.g. in Ref.\cite{LS2012}.

The corresponding differential cross section corresponding
to this contribution can be calculated as:
\begin{equation}
\frac{d\sigma}{d y_+ d y_- d^2 p_{W\perp}} = \frac{1}{16 \pi^2 {\hat s}^2}
\, x_1 g(x_1,\mu^2) \, x_2 g(x_2,\mu^2)
\overline{ | {\cal M}_{g g \to W^+ W^-}(\hat s, \hat t, \hat u)
  |^2} \, ,
\label{gg_fusion}
\end{equation}
where g's are gluon distributions in the proton.
This contribution is formally higher order in pQCD than 
the $q \bar q$ annihilation, but may be large numerically at higher energies
when $x_1$ and $x_2$ become very small, i.e. gluon distributions
are very large.

\subsubsection{$pp \to W^+ W^-X$ production via photon exchanges}

In this section, we briefly discuss the inclusive $pp \to W^+ W^-X$ 
mechanism via photon exchanges where we consider the cases when the protons are
destroyed in the final state (inelastic contributions)
thus leading to a $W$ pair and in addition the proton remnants. 
We calculate this contribution to the inclusive
$p p \to W^+ W^- X$ process for the first time in the litterature.

\begin{figure*}
\begin{center}
\includegraphics[width=5cm,height=4cm]{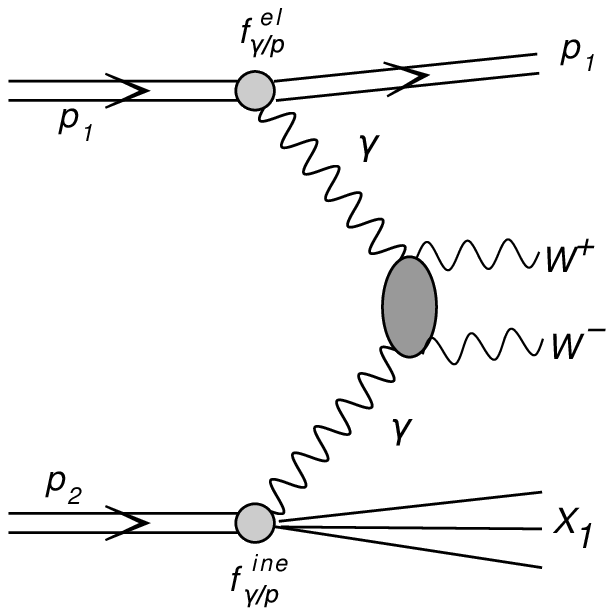}
\includegraphics[width=5cm,height=4cm]{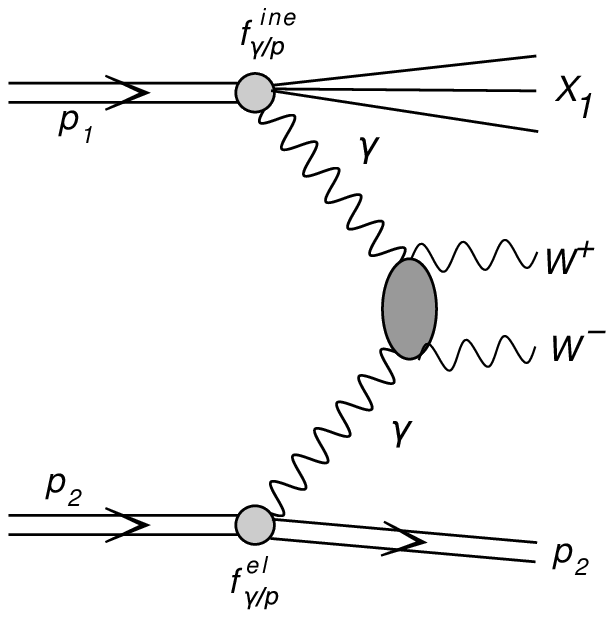}
\includegraphics[width=5cm,height=4cm]{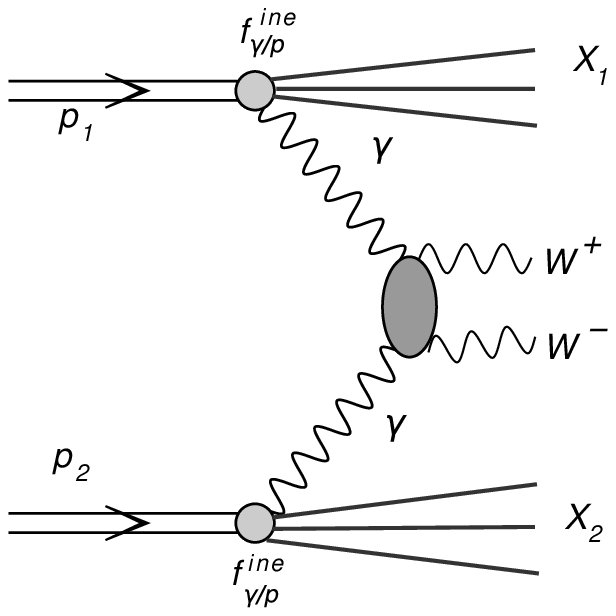}
\caption{Diagrams representing inelastic photon-photon induced mechanisms 
for production of $W^+ W^-$ pairs.
}
\label{fig:new_diagrams}
\end{center}
\end{figure*}

If at least one photon is a constituent of the nucleon, 
the mechanisms presented in Fig.\ref{fig:new_diagrams} are possible.
In these cases at least one of participating protons does not survive 
the $W^+ W^-$ production process. In the following we consider 
two different approaches to the problem.
\\

{\bf Naive approach to photon flux}
\\

Some $\gamma \gamma$ induced processes 
($\gamma \gamma \to H^+ H^-, L^+ L^-$) were discussed long time ago
in Ref.\cite{DGNR94}. In these approaches the photon distribution
in the proton is a convolution of the distribution of quarks
in the proton and the distribution of photons in quarks/antiquarks
\begin{equation}
f_{\gamma/p} = f_q \otimes f_{\gamma/q}   \; ,
\end{equation}
which can be written mathematically as
\begin{equation}
x f_{\gamma/p}(x) = \sum_{q} \int_x^1 d x_q f_q(x_q,\mu^2) 
e_q^2 \left( \frac{x}{x_q} \right) f_{\gamma/q}
\left( \frac{x}{x_q},Q_1^2,Q_2^2 \right) \; ,
\label{convolution}
\end{equation}
where the sum runs over all quark and antiquark flavours.
The flux of photons in a quark/antiquark in their approach 
was calculated as:
\begin{equation}
f_{\gamma}(z) = \frac{\alpha_{em}}{2 \pi}
\frac{1 + (1-z)^2}{2} \log \left( \frac{Q_1^2}{Q_2^2} \right) \; .
\label{photon_in_quark}
\end{equation}
The choice of the scales in the formulae is a bit ambigous.
In these papers the authors have proposed the following set of scales:
\begin{eqnarray}
Q_1^2 &=& \max ({\hat s}/4-m_W^2, 1 \mbox{GeV}^2) \nonumber \\
Q_2^2 &=& 1 \mbox{GeV}^2 \nonumber \\
\mu^2 &=& {\hat s}/4 \; .
\label{scales}
\end{eqnarray}
We use the approach decribed in this section as a reference 
for the formally more refined calculation described in the next subsection.
\\

{\bf MRST-QED parton distributions}
\\

An improved approach how to include photons into inelastic processes 
was proposed some time ago by Martin, Roberts, 
Stirling and Thorne in Ref.\cite{MRST04}. In their approach the photon 
is treated on the same footing as quarks, antiquarks and gluons.
Below we repeat the essential points of their formalism which includes 
combined QCD+QED evolution.

They proposed a QED-corrected evolution equations for the parton 
distributions of the proton \cite{MRST04}:
\begin{eqnarray}
{\partial q_i(x,\mu^2) \over \partial \log \mu^2} &=& {\alpha_S\over 2\pi}
\int_x^1 \frac{dy}{y} \Big\{
    P_{q q}(y)\; q_i(\frac{x}{y},\mu^2)
     +  P_{q g}(y)\; g(\frac{x}{y},\mu^2)\Big\}
\, \nonumber \\
&  + &
   {\alpha\over 2\pi} \int_x^1 \frac{dy}{y} \Big\{
    \tilde{P}_{q q}(y)\; e_i^2 q_i(\frac{x}{y},\mu^2)  +  P_{q \gamma}(y)\;
e_i^2 \gamma(\frac{x}{y},\mu^2)         \Big\},  \nonumber \\
{\partial g(x,\mu^2) \over \partial \log \mu^2} &=& {\alpha_S\over 2
\pi} \int_x^1 \frac{dy}{y} \Big\{
    P_{g q}(y)\; \sum_j q_j(\frac{x}{y},\mu^2) 
 + 
    P_{g g}(y)\; g(\frac{x}{y},\mu^2)\Big\},
\, \nonumber \\
   {\partial \gamma(x,\mu^2) \over \partial \log \mu^2}
& =   & {\alpha
\over 2\pi} \int_x^1 \frac{dy}{y} 
   \Big\{ P_{\gamma q}(y)\; \sum_j e_j^2\; q_j(\frac{x}{y},\mu^2) 
+ 
  P_{\gamma \gamma}(y)\; \gamma(\frac{x}{y},\mu^2) \Big\} \; ,
\label{evolution_equations}
\end{eqnarray}
where
\begin{eqnarray}
{\tilde P}_{qq} = C_F^{-1} P_{qq}, & &   P_{\gamma q} = 
C_F^{-1} P_{g q}, \nonumber \\
P_{q\gamma} = T_R^{-1} P_{q g} , & &  P_{\gamma \gamma} = - 
\frac{2}{3}\; \sum_i e_i^2\; \delta(1-y) \nonumber
\end{eqnarray}
The parton distributions in Eq.(\ref{evolution_equations}) fulfil 
the standard momentum sum rule:
\begin{equation}
  \int_0^1 dx\;  x\; \Big\{\sum_i q_i(x,\mu^2) + g(x,\mu^2) + \gamma(x,\mu^2)
     \Big\}  = 1 \; .
\end{equation}
\\

{\bf Cross section for photon-photon processes}
\\

At leading order, the corresponding triple differential cross section 
for inelastic-inelastic photon-photon contribution can be written as
usually in the parton-model formalism:
\begin{eqnarray}
\frac{d \sigma^{\gamma_{in} \gamma_{in}}}{d y_1 d y_2 d^2p_t} &=& \frac{1}{16 \pi^2 {\hat s}^2}
x_1 \gamma_{in}(x_1,\mu^2) \; x_2 \gamma_{in}(x_2,\mu^2) \;
\overline{|{\cal M}_{\gamma \gamma \to W^+W^-}|^2} \; .
\end{eqnarray}
Above $\gamma_{in}(...)$ is used to denote "inelastic" photon distribution in
the proton ($\gamma(...)$ in the previous section).
$\gamma_{el}(...)$ will be reserved for purely elastic cases when proton stays intact.

The above contribution includes only cases when both nucleons do not survive
the collision and nucleon debris are produced. The case when
at least one nucleon survives the collision has to be considered
separately. In this case one can include the corresponding photon 
distributions where an extra "el" index will be added. 
The corresponding contributions to the cross section 
can be then written as:
\begin{eqnarray}
\frac{d \sigma^{\gamma_{in} \gamma_{el}}}{d y_1 d y_2 d^2p_t} &=& \frac{1}{16 \pi^2 {\hat s}^2}
x_1 \gamma_{in}(x_1,\mu^2) \; x_2 \gamma_{el}(x_2,\mu^2) \;
\overline{|{\cal M}_{\gamma \gamma \to W^+W^-}|^2} \; ,\nonumber \\
\frac{d \sigma^{\gamma_{el} \gamma_{in}}}{d y_1 d y_2 d^2p_t} &=& \frac{1}{16 \pi^2 {\hat s}^2}
x_1 \gamma_{el}(x_1,\mu^2) \; x_2 \gamma_{in}(x_2,\mu^2) \;
\overline{|{\cal M}_{\gamma \gamma \to W^+W^-}|^2} \; ,\nonumber \\
\frac{d \sigma^{\gamma_{el} \gamma_{el}}}{d y_1 d y_2 d^2p_t} &=& \frac{1}{16 \pi^2 {\hat s}^2}
x_1 \gamma_{el}(x_1,\mu^2) \; x_2 \gamma_{el}(x_2,\mu^2) \; 
\overline{|{\cal M}_{\gamma \gamma \to W^+W^-}|^2} \; , \\ 
\label{subleading_contributions}
\end{eqnarray}
for inelastic-elastic, elastic-inelastic and elastic-elastic
components, respectively.
The last case was already discussed in a separate dedicated subsection.
In the following the elastic photon fluxes are calculated using 
the Drees-Zeppenfeld parametrization \cite{DZ}, where a simple 
parametrization of the nucleon electromagnetic form factors is used.
Such an approach is consistent with the partonic approach used
recently to estimate the electroweak corrections to different QCD processes.
The three terms shown above are usually omitted.
In this paper we quantify these contributions for $W^+ W^-$ production.
\\

{\bf Resolved photons}
\\

\begin{figure*}
\begin{center}
\includegraphics[width=4cm,height=5cm]{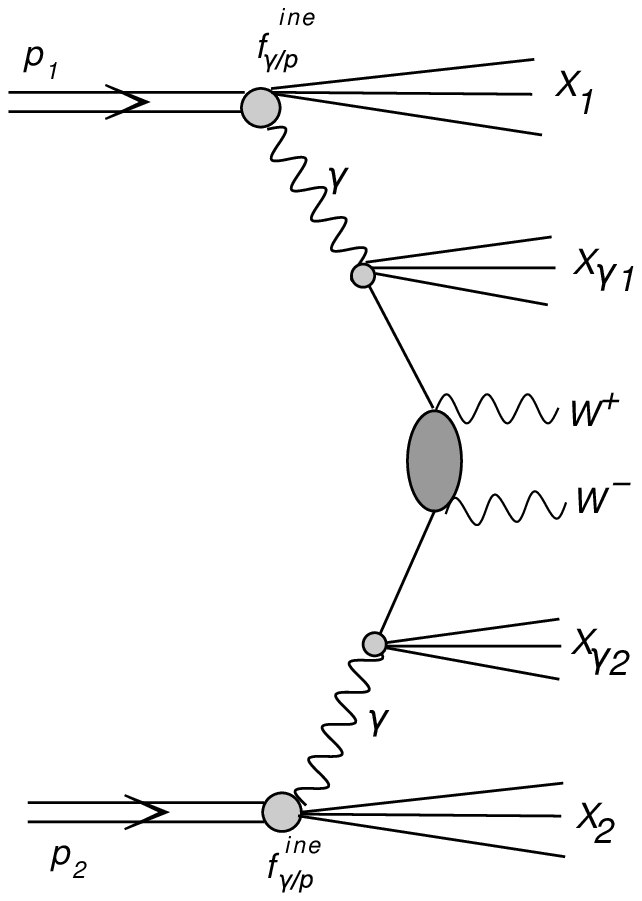}
\includegraphics[width=4cm,height=5cm]{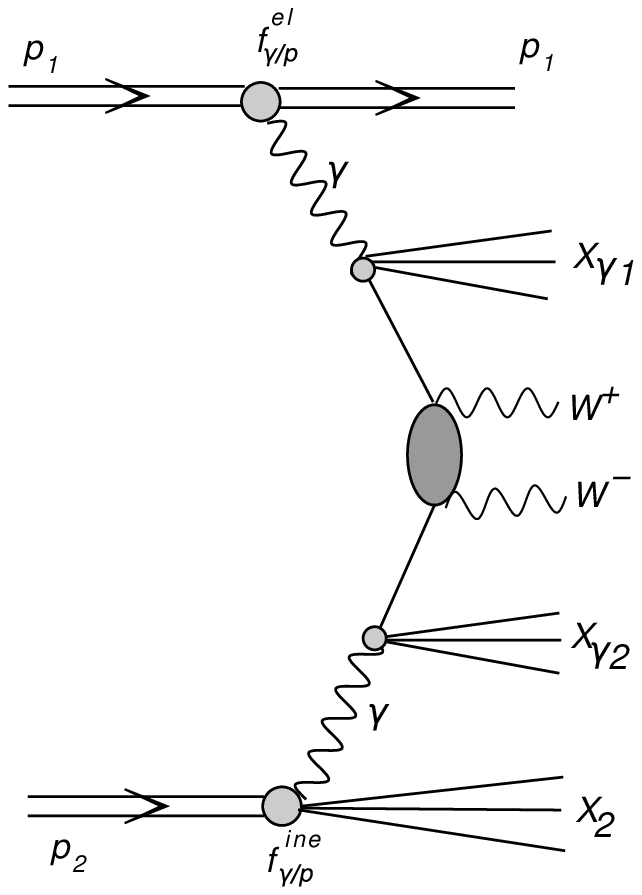}
\includegraphics[width=4cm,height=5cm]{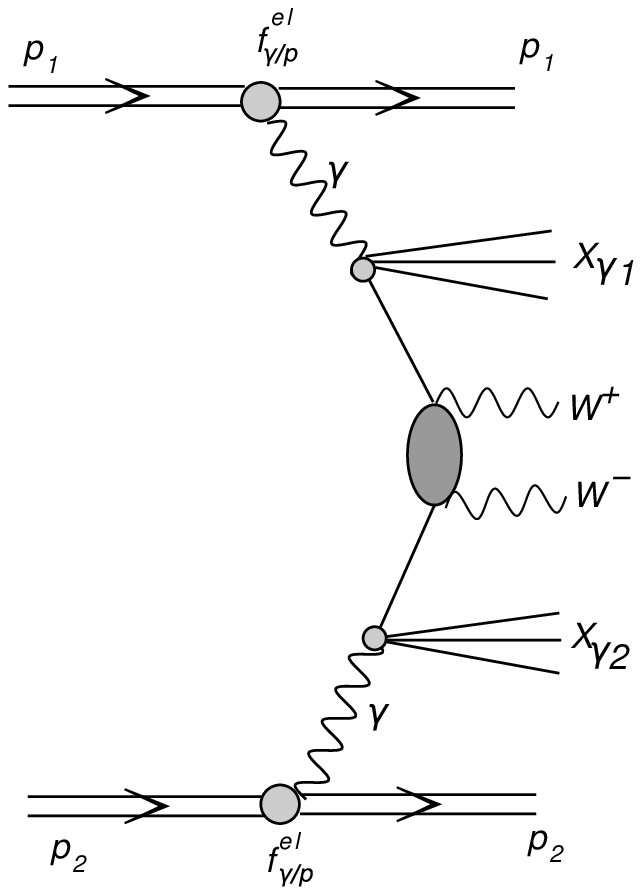}
\includegraphics[width=4cm,height=5cm]{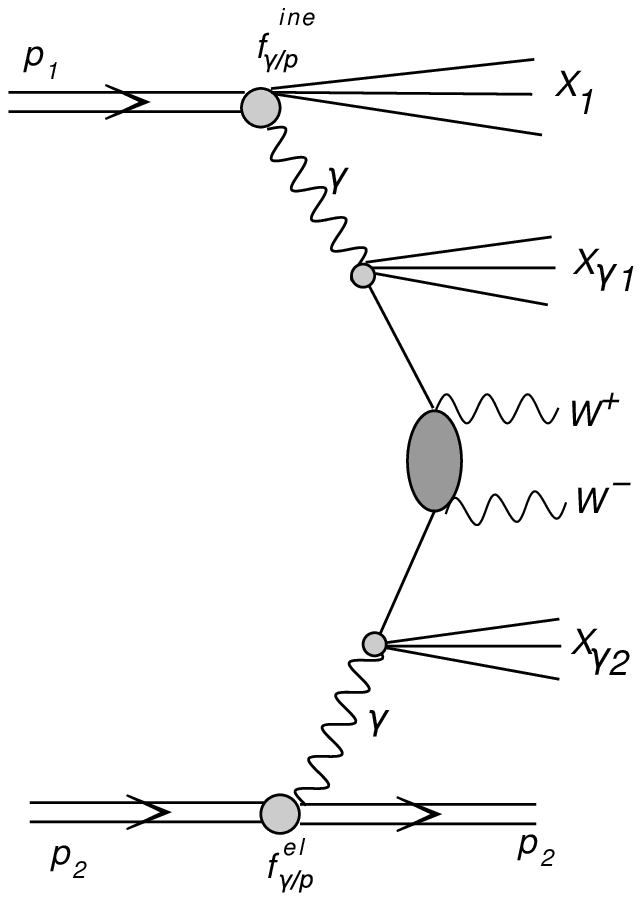}
\caption{Diagrams representing some single resolved photon 
mechanisms with quark-antiquark annihilation for production of $W^+ W^-$
pairs. Similar diagrams with gluon-gluon subprocesses also exist.
}
\label{fig:resolved1}
\end{center}
\end{figure*}

So far we have discussed direct photonic contributions. We also need to add
the hadronic content of the photon. Asymmetric diagrams with one photon
attached to the upper or lower proton,
as shown in Fig.\ref{fig:resolved1}, become possible.
Extra photon remnant debris 
(called $X_{\gamma, 1}$ or $X_{\gamma, 2}$ in the figure) 
appear in addition. One may expect that such diagrams lead to quite
asymmetric distributions in $W$ boson rapidity with maxima in 
forward and/or backward directions.

\begin{figure*}
\begin{center}
\includegraphics[width=4cm,height=5cm]{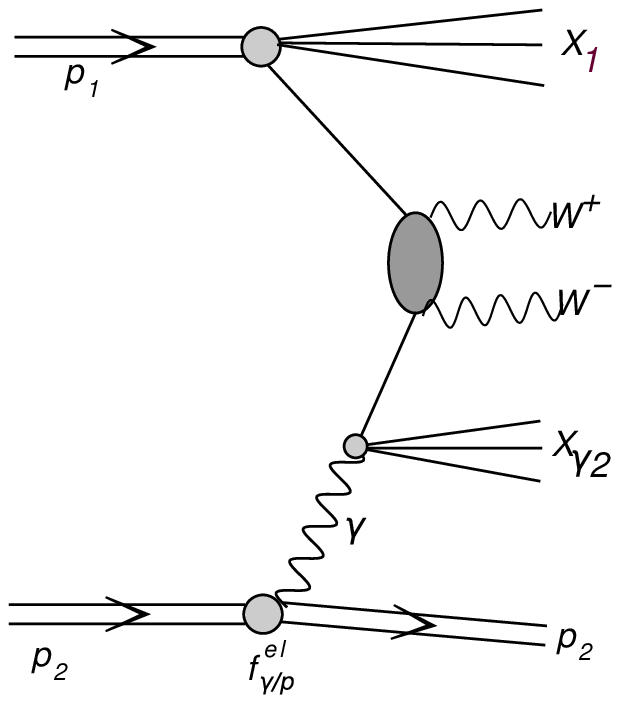}
\includegraphics[width=4cm,height=5cm]{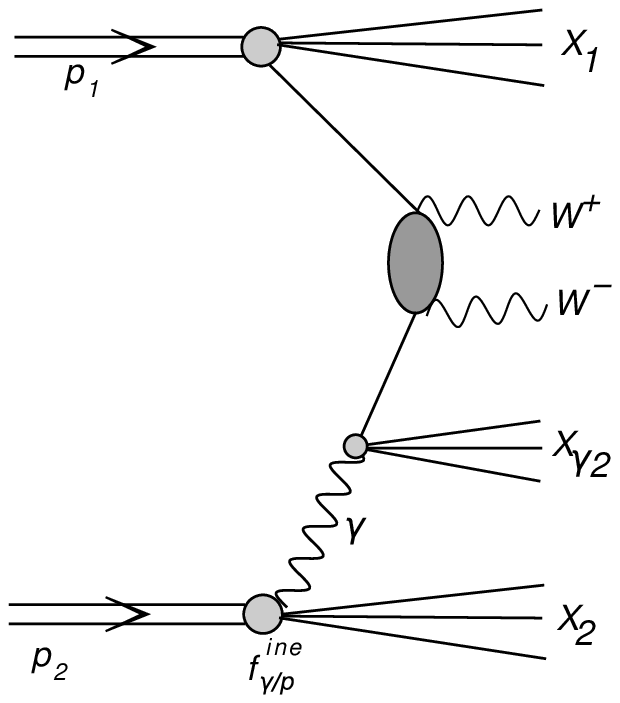}
\includegraphics[width=4cm,height=5cm]{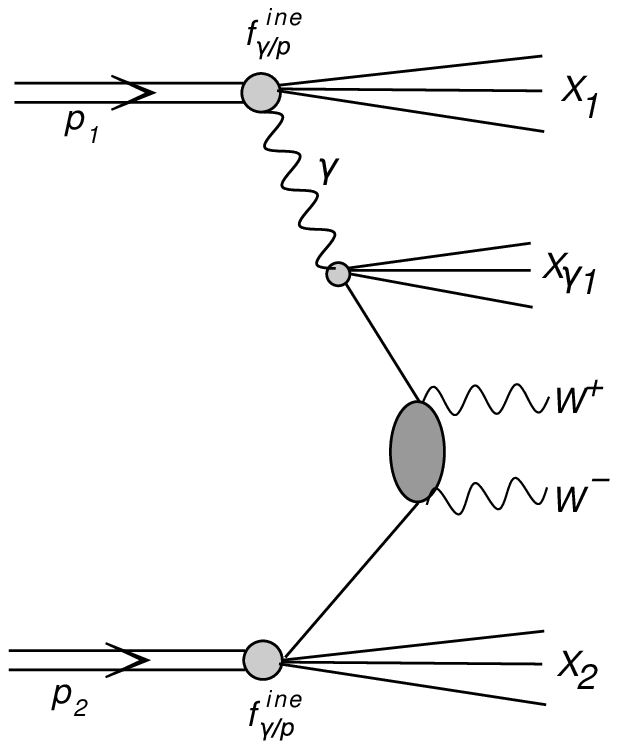}
\includegraphics[width=4cm,height=5cm]{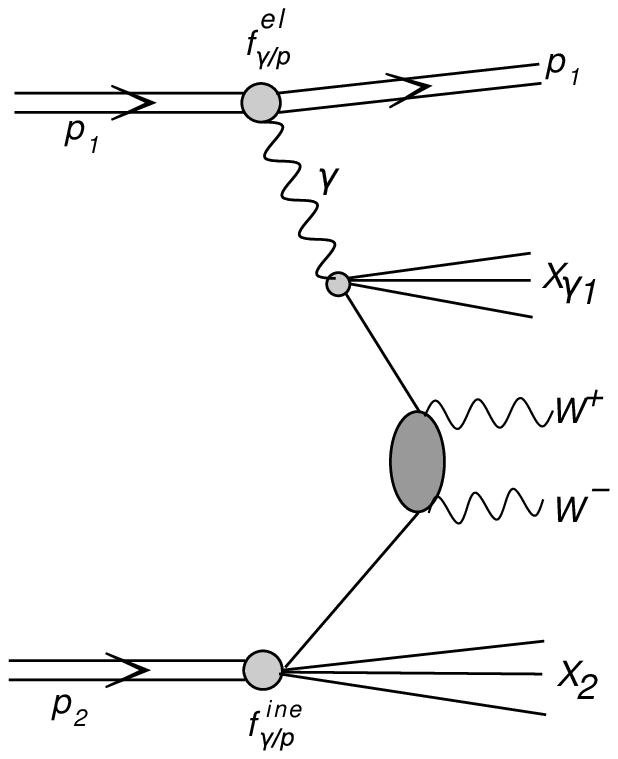}
\caption{Diagrams representing mechanisms with two resolved photons 
for the production of $W^+ W^-$ pairs.
}
\label{fig:resolved2}
\end{center}
\end{figure*}

Another type of diagrams with resolved photons is shown in
Fig.\ref{fig:resolved2}. We expect the contributions of the second
set of diagrams to be rather
small. 

In the case of resolved photons, the ``photonic'' quark/antiquark 
distributions in a proton must be calculated first. This can be done 
as the convolution
\begin{equation}
f_{q/p}^{\gamma} = f_{\gamma/p} \otimes f_{q/\gamma}
\end{equation}
which mathematically means:
\begin{equation}
x f_{q/p}^{\gamma}(x) = \int_x^1 d x_{\gamma} f_{\gamma/p}(x_{\gamma},\mu_s^2) 
\left( \frac{x}{x_{\gamma}} \right) 
f \left( \frac{x}{x_{\gamma}}, \mu_h^2 \right)  \; . 
\label{convolution_resolved}       
\end{equation}
Technically the first $f_{\gamma/p}$ in the proton is computed on a dense
grid for $\mu_s^2 \sim$ 1 GeV$^2$ (virtuality of the photon) and  
it is used in the convolution formula (\ref{convolution_resolved}). 
The second scale is evidently hard $\mu_h^2 \sim M_{WW}^2$.
The result strongly depends on the choice
of the soft scale $\mu_s^2$. In this sense our calculations
are not very precise and must be treated rather as a rough estimate.
The new quark/antiquark distributions of photonic origin are 
used to calculate the cross section as for the standard quark-antiquark 
annihilation subprocess.

\subsubsection{Single diffractive production of $W^+ W^-$ pairs}

Diffractive processes for $W^+ W^-$ production were not considered so
far in the litterature but are potentially very important.

\begin{figure*}
\begin{center}
\includegraphics[width=5cm,height=5cm]{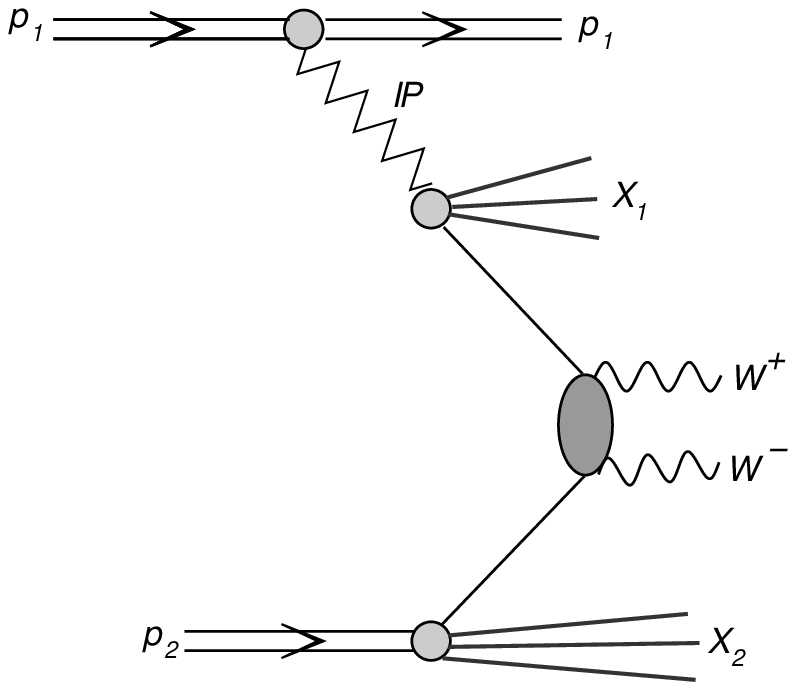}
\includegraphics[width=5cm,height=5cm]{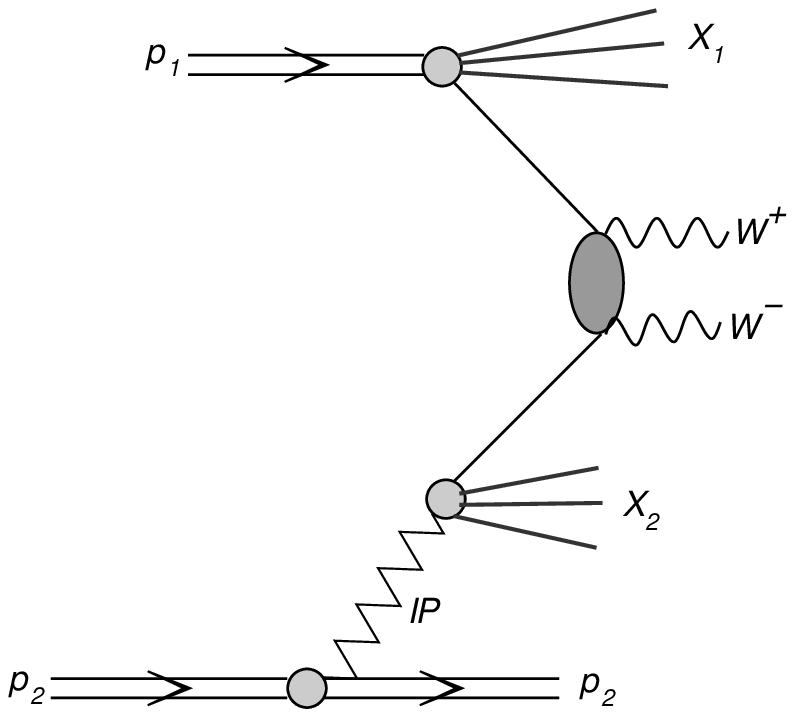}
\caption{Diagrams representing single diffractive mechanism
of production of $W^+ W^-$ pairs.
}
\label{fig:single_diffractive}
\end{center}
\end{figure*}

In the following we apply the resolved pomeron approach 
\cite{IS_ee,IS_ccbar}.
In this approach one assumes that the Pomeron has a
well defined partonic structure, and that the hard process
takes place in Pomeron--proton or proton--Pomeron (single diffraction) 
or Pomeron--Pomeron (central diffraction) processes.
The mechanism of single diffractive production of $W^+ W^-$ pairs
is shown in Fig.\ref{fig:single_diffractive}.
We calculate the triple differential distributions
\begin{eqnarray}
{d \sigma_{SD}^{(1)} \over dy_{1} dy_{2} dp_{t}^2} = \sum_{f}{\Big| M \Big|^2 \over 16 \pi^2 \hat{s}^2} 
\,\Big [\, \Big( x_1 q_f^D(x_1,\mu^2) 
\, x_2 \bar q_f(x_2,\mu^2) \Big) \, 
+ \Big( x_1 \bar q_f^D(x_1,\mu^2)
\, x_2  q_f(x_2,\mu^2) \Big) \, \Big ] ,
\nonumber \\ 
\label{DY}
\end{eqnarray}
\begin{eqnarray}
{d \sigma_{SD}^{(2)} \over dy_{1} dy_{2} dp_{t}^2} = \sum_{f}{\Big| M \Big|^2 \over 16 \pi^2 \hat{s}^2} 
\,\Big [\, \Big( x_1 q_f(x_1,\mu^2) 
\, x_2 \bar q_f^D(x_2,\mu^2) \Big) \, 
+ \Big( x_1 \bar q_f(x_1,\mu^2)
\, x_2  q_f^D(x_2,\mu^2) \Big) \, \Big ] ,
\nonumber \\ 
\label{SD}
\end{eqnarray}
\begin{eqnarray}
{d \sigma_{CD} \over dy_{1} dy_{2} dp_{t}^2} =  \sum_{f}{\Big| M \Big|^2 \over 16 \pi^2 \hat{s}^2} 
\,\Big [\, \Big( x_1 q_f^D(x_1,\mu^2) 
\, x_2 \bar q_f^D(x_2,\mu^2) \Big) \, 
+ \Big( x_1 \bar q_f^D(x_1,\mu^2)
\, x_2  q_f^D(x_2,\mu^2) \Big) \,\Big ] 
\nonumber \\ 
\label{DD}
\end{eqnarray}
for single-diffractive and central-diffractive production, 
respectively.
$q_f^D$ and $\bar q_f^D$ are so - called diffractive quark and antiquark
distributions, respectively. We shall return to them in the framework of the 
resolved pomeron model somewhat below.
The matrix element squared for the $q \bar q \to W^{+} W^{-}$
process is the same as previously discussed for the non-diffractive processes.

In this approach longitudinal momentum fractions are calculated as
\begin{eqnarray}
x_1 = {m_{t} \over \sqrt{s}}  \Big( e^{y_{2}} + e^{y_{2}} \Big) , \\
x_2 = {m_{t} \over \sqrt{s}}  \Big( e^{-y_{1}} + e^{-y_{2}} \Big)
\nonumber
\end{eqnarray}
with $m_{t} = \sqrt{ (p_{t}^2 + m_{W}^2)} \approx p_{t} $.
The distribution in the $WW$ invariant mass can be 
obtained by binning differential cross section in $M_{WW}$.

In the present analysis we consequently do not calculate higher-order 
contributions. In principle, they could be included effectively with 
the help of a so-called $K$-factor.

The ``diffractive" quark/antiquark distribution of
flavour $f$ can be obtained by a convolution of the Pomeron flux
$f_\Pom(x_\Pom)$ and the parton distribution in the Pomeron 
$q_{f/\Pom}(\beta, \mu^2)$:
\begin{eqnarray}
q_f^D(x,\mu^2) = \int d x_\Pom d\beta \, \delta(x-x_\Pom \beta) 
q_{f/\Pom} (\beta,\mu^2) \, f_\Pom(x_\Pom) \, 
= \int_x^1 {d x_\Pom \over x_\Pom} \, f_\Pom(x_\Pom)  
q_{f/\Pom}({x \over x_\Pom}, \mu^2) \, . \nonumber \\
\label{diffractive_convolution}
\end{eqnarray}
The Pomeron flux $f_\Pom(x_\Pom)$ is integrated over 
the four--momentum transfer 
\begin{eqnarray}
f_\Pom(x_\Pom) = \int_{t_{min}}^{t_{max}} dt \, f(x_\Pom,t) \, ,
\label{flux_of_Pom}
\end{eqnarray}
with $t_{min}, t_{max}$ being kinematic boundaries.

Both pomeron flux factors $f_{\Pom}(x_{\Pom},t)$ and the
quark/antiquark distributions in the pomeron are taken from 
the H1 collaboration analysis of diffractive structure function
and diffractive dijets at HERA \cite{H1}. 
The factorization scale for diffractive parton distributions is taken 
here as $\mu^2 = m_t^2$.

In the present analysis we consider both pomeron and subleading reggeon
contributions. In the H1 collaboration analysis the pion structure 
function was used for the subleading reggeons and the corresponding 
flux was fitted to the diffractive DIS data.
The corresponding diffractive quark distributions
are obtained by replacing the pomeron flux by the reggeon flux and
the quark/antiquark distributions in the pomeron by their counterparts
in subleading reggeon(s). Additional details can be found in \cite{H1}.
In the case of pomeron exchange the upper limit in 
(\ref{diffractive_convolution}) is taken to be 0.1 and for reggeon exchange 0.2.
In our opinion, the whole Regge formalism does not apply above these limits
and therefore unphysical results could be obtained.

Up to now we have assumed Regge factorization which is known
to be violated in hadron-hadron collisions.
It is known that these are soft interactions which lead to an extra 
production of particles which fill in the rapidity gaps related 
to pomeron exchange.

If rapidity gap (gaps) is required (measured) then one has to include
absorption effect in the formalism of the resolved pomeron/reggeon
which can be interpreted as a probability of no extra soft interactions
leading to the destruction of the rapidity gap.

Different models of absorption corrections 
(one-, two- or three-channel approaches) 
for diffractive processes were presented in the litterature.
The absorption effects for diffractive processes were calculated e.g.
in Ref \cite{Khoze,Maor}.
The different models give slightly different predictions.
Usually an average value of the gap survival probability
$<|S|^2>$ is calculated first and the cross sections for different
processes are multiplied by this value.
We shall follow this somewhat simplified approach.
Numerical values of the gap survival probability can be found 
in \cite{Khoze,Maor}.
The survival probability depends on the collision energy.
It is sometimes parametrized as:
\begin{equation}
<|S|^2>(\sqrt{s}) = \frac{a}{b+\ln(\sqrt{s})} \; .
\end{equation}
The numerical values of the parameters can be found in original
publications. At the LHC energy of 14 TeV one gets typically $S^2_G$ = 0.03.
The diffractive cross sections at 8 TeV below will be multiplied by 
the gap survival factor $S^2_G$ = 0.08 extracted recently by the CMS collaboration
\cite{CMS_2013} for diffractive dijet production.
In our opinion
there is about 30 \% uncertainty on this value and it will be important to
measure it using the incoming data at 13 TeV.
In general, the absorptive corrections for single and 
central diffractive processes could be somewhat different.

\subsubsection{Double parton scattering}

\begin{figure*}
\begin{center}
\includegraphics[width=5cm,height=5cm]{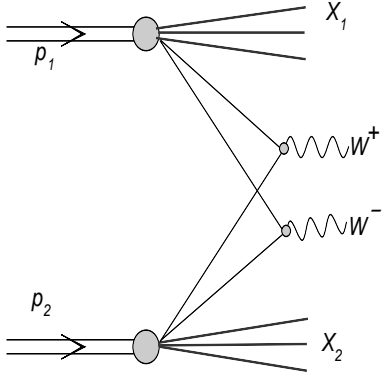}
\caption{Diagram representing double parton scattering mechanism
of production of $W^+ W^-$ pairs.
}
\label{fig:DPS}
\end{center}
\end{figure*}

For completeness we also consider the double parton scattering
mechanism discussed already in the litterature but not included in the
Standard Model predictions to the inclusive cross section
and distributions.
The diagram representating the double parton scattering
process is shown in Fig.\ref{fig:DPS}.

The cross section for double parton scattering is often modelled
in the factorized anzatz which in our case would mean:
\begin{equation}
\sigma_{W^+ W^-}^{DPS} = \frac{1}{\sigma_{qq}^{eff}} 
\sigma_{W^{+}}
\sigma_{W^{-}}
\; .
\label{factorized_model}
\end{equation}
This is a rather effective approach assuming implicitly the lack of parton
correlations. In principle also single parton splitting mechanisms
should be included explicitly (see e.g. \cite{Gaunt}).
The phenomenological formula includes them effectively.

In general, the parameter $\sigma_{q q}$ does not need to be the same
as for gluon-gluon initiated processes, which is better known 
phenomenologically ($\sigma_{gg}^{eff} \approx$ 15 mb) from analyses 
of different experimental data.
In the present, rather conservative, calculations we take it to be
$\sigma_{qq}^{eff} = \sigma_{gg}^{eff}$ = 15 mb.
The latter value is known within about 10 \% from 
systematics of gluon-gluon initiated processes at the Tevatron and LHC.

The factorized model (\ref{factorized_model}) can be generalized 
to more differential distributions (see e.g. \cite{LMS2012,MS2013}).
For example in our case of $W^{+} W^{-}$ production the cross section
differential in $W$ boson rapidities can be written as:
\begin{equation}
\frac{d \sigma_{W^+ W^-}^{DPS}}{d y_{+} d y_{-}} =
\frac{1}{\sigma_{qq}^{eff}} 
\frac{d\sigma_W^{+}}{d y_{+}}
\frac{d\sigma_W^{-}}{d y_{-}} \; '
\label{generalized_factorized_model}
\end{equation}
where $y_{+}$ and $y_{-}$ are rapidites of $W^+$ and $W^-$, respectively.
In particular, in leading-order approximation the cross section for 
quark-antiquark annihilation reads:
\begin{equation}
\frac{d\sigma}{dy} = \sum_{ij} 
\left( 
  x_1 q_{i/1}(x_1,\mu^2) x_2 {\bar q}_{j/2}(x_2,\mu^2) 
+ x_1 {\bar q}_{i/1}(x_1,\mu^2) x_2 q_{j/2}(x_1,\mu^2) \right)
\overline{|{\cal M}_{ij \to W^{\pm}}|^2} \; ,
\label{rapidity_of_W}
\end{equation}
where the matrix element for quark-antiquark annihilation to $W$ bosons
(${\cal M}_{ij \to W^{\pm}}$) contains the Cabibbo-Kobayashi-Maskawa 
matrix elements.
In the present paper we show the 
$\frac{d\sigma}{dy_{+} dy_{-}}$ cross section, as well as the distributions
in rapidity difference between $W^{+}$ and $W^{-}$ and 
in $M_{WW}$.
In the approximations made here (leading order approximation, no
transverse momenta of W bosons)
\begin{equation}
M_{WW}^2 = 2 M_W^2 \left( 1 + \mbox{cosh}(y_1-y_2) \right) \; .
\label{invariant_mass}
\end{equation}

When calculating the cross section for single $W$ boson production
in leading-order approximation a well known Drell-Yan $K$-factor could 
be included. The double-parton scattering would be then multiplied by
$K^2$. The $K$-factor for  $W^{+}$ or $W^{-}$ production is only slightly
larger then 1 and in the consequence also $K^2$ is not far from 1.

Fast progress in understanding multi-parton interactions was achieved 
recently. It is clear at present that in addition to the ususal DPS
terms, the perturbative parton splitting mechanism has to be included
explicitly \cite{RS2011,BDFS2013}.
The contribution of this mechanism can be as large as the conventional
DPS contribution. This formalism contains, however, more
partonic form factors that are not well known phenomenologically.
We think that the estimate of the DPS effect based on the factorized
Ansatz with empirical $\sigma_{eff}$ can be better, at least
in the moment. Further studies are needed also here.
It is not easy to estimate the uncertainty of the so calculate DPS 
cross sections but we expect that the order of magnitude should be correct.

\section{Results}
\label{results}

In this section, we evaluate each contribution decribed in Section II
and try to see if they can be measured experimentally in some dedicated kinematical
regions.

\subsection{Discussion of the photon induced inelastic contributions}

\begin{figure}
\begin{center}
\includegraphics[width=8cm,height=5cm]{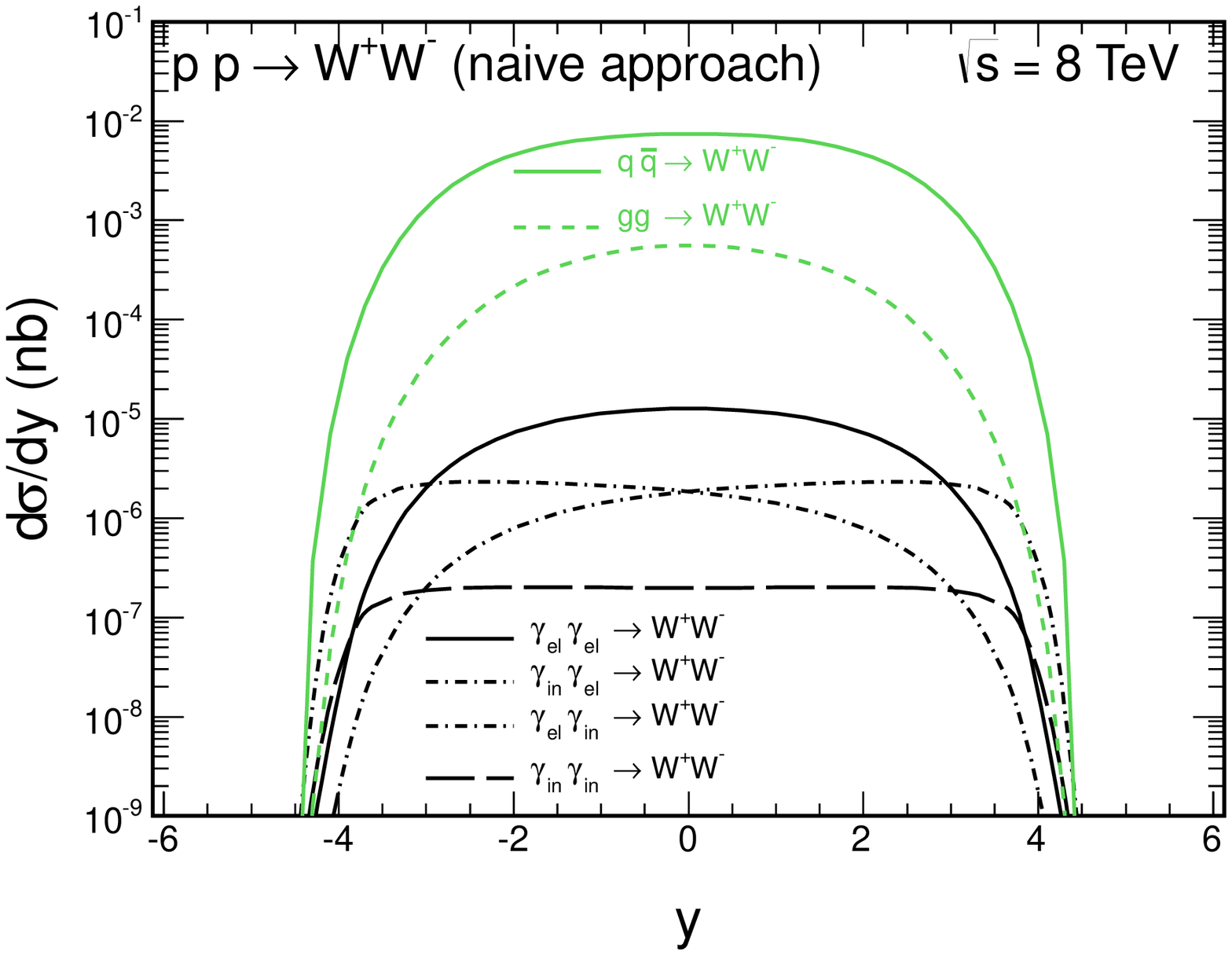}
\includegraphics[width=8cm,height=5cm]{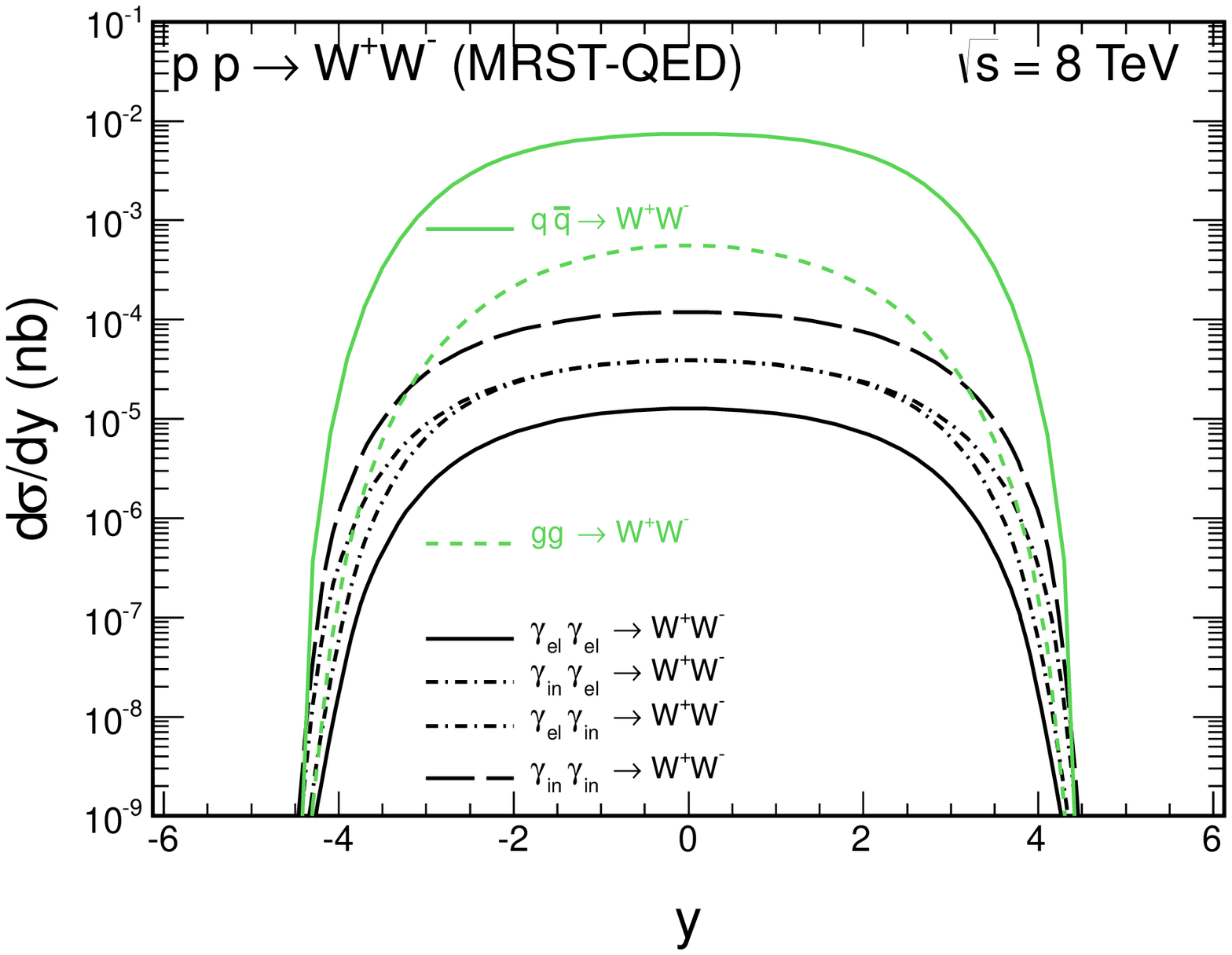}
\end{center}
\caption {Rapidity distribution of $W$ bosons for $\sqrt{s}$ = 8 TeV.
The left panel shows the results for the naive approach often used
in the litterature, while the right panel shows the result with
the QCD improved method proposed in Ref.\cite{MRST04}. 
}
\label{fig:dsig_dy_approx1}
\end{figure}

Before a detailed survey of results of the different contributions discussed
in the present paper, let us concentrate on
some technical details concerning inelastic photon-photon contributions.
In Fig.\ref{fig:dsig_dy_approx1} we show the rapidity distributions for 
the naive (left panel) and QCD improved (right panel) approaches
discussed in section II. 
While in the naive approximation the elastic-elastic component
is the largest and inelastic-inelastic is the smallest, in 
the QCD improved approach the situation is reversed, which shows the importance
to use the complete approach and not the naive one. Here, in the QCD 
improved calculations $\mu_F^2 = m_t^2$ was used as 
the factorization scale.

\begin{figure}
\begin{center}
\includegraphics[width=8cm,height=5cm]{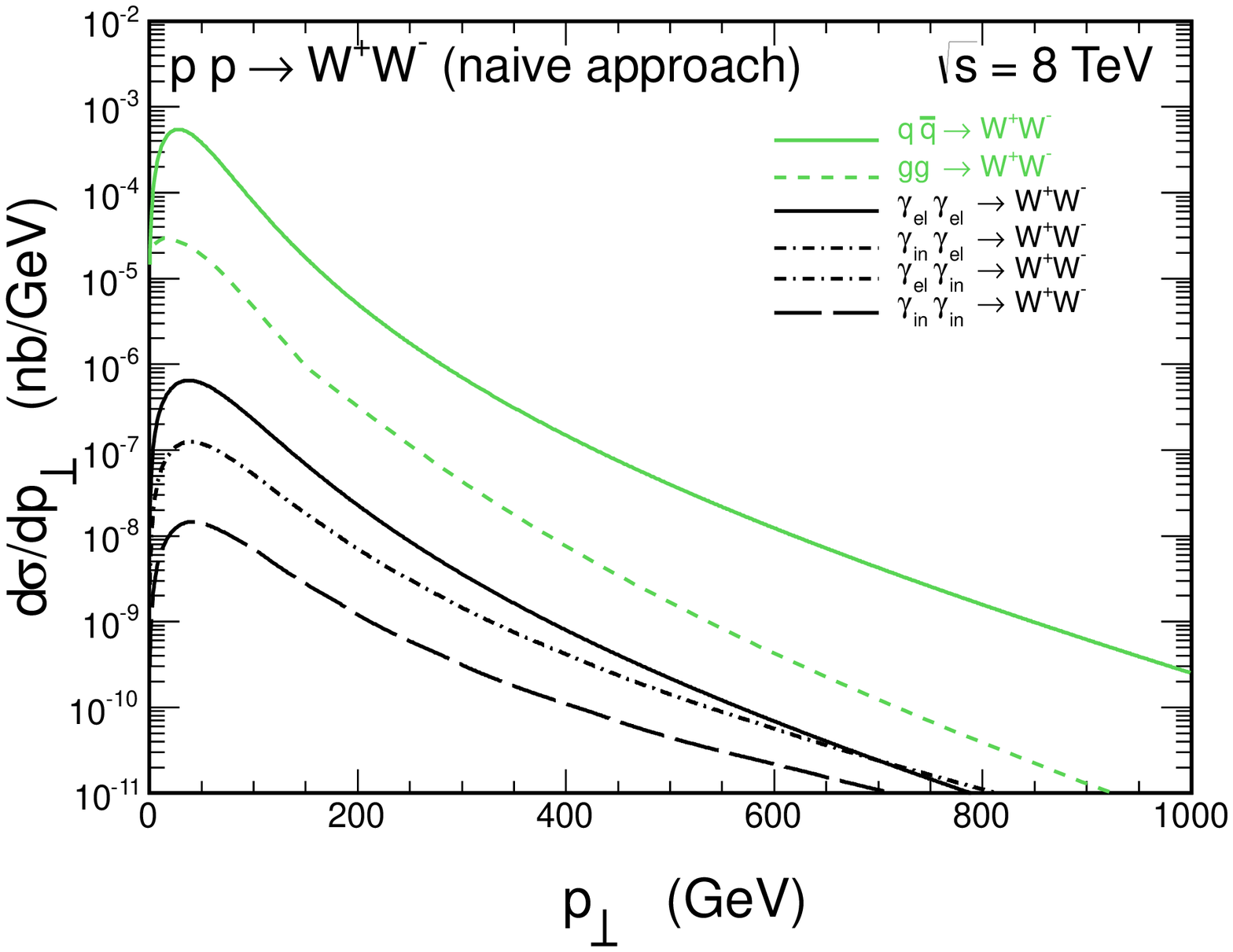}
\includegraphics[width=8cm,height=5cm]{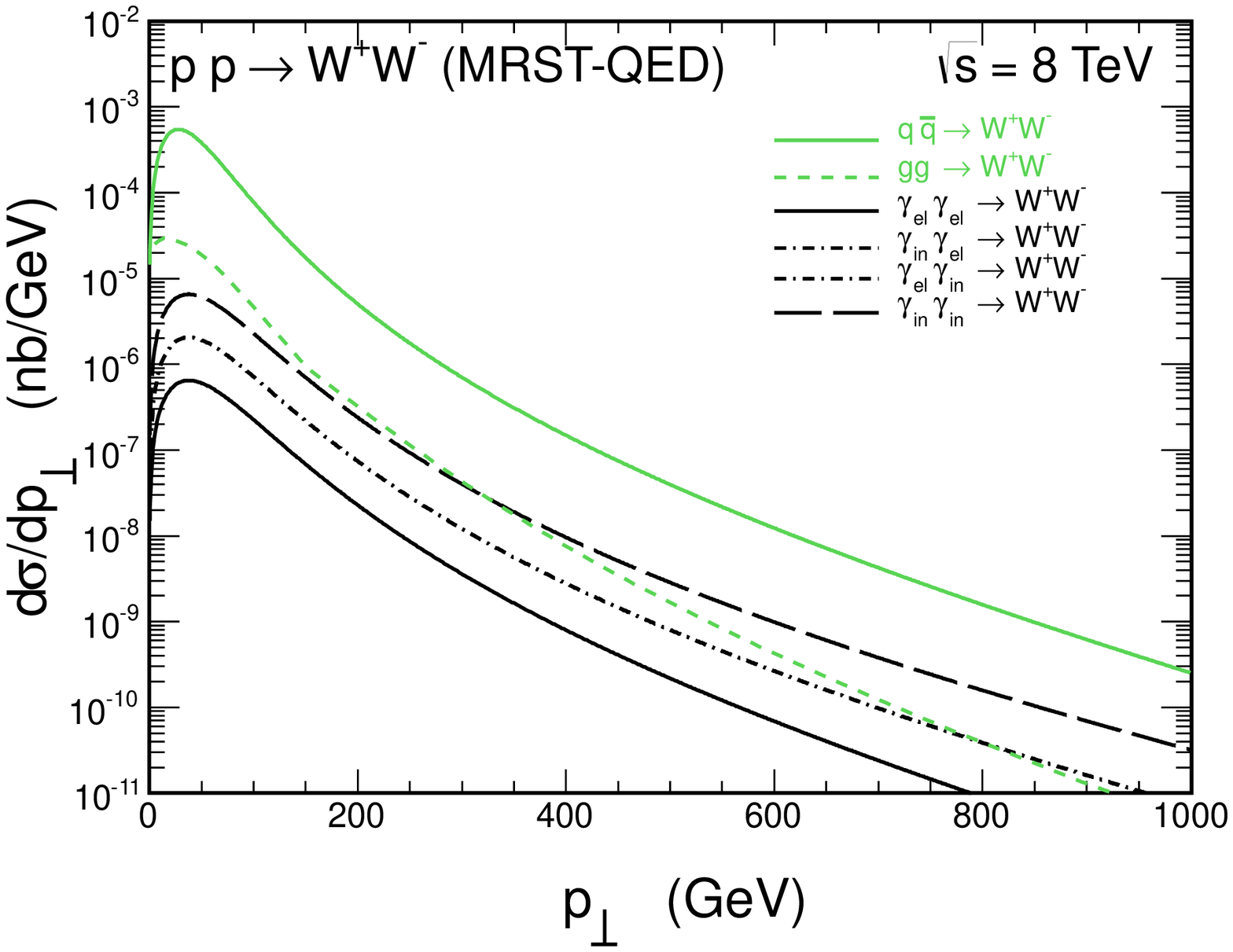}
\end{center}
\caption {Transverse momentum distribution of $W$ bosons for 
$\sqrt{s}$ = 8 TeV.
The left panel shows results for the naive approach often used
in the past, while the right panel shows the result with
the QCD improved method proposed in Ref.\cite{MRST04}.
}
\label{fig:dsig_dy_approx}
\end{figure}

\begin{figure}
\begin{center}
\includegraphics[width=8cm,height=5cm]{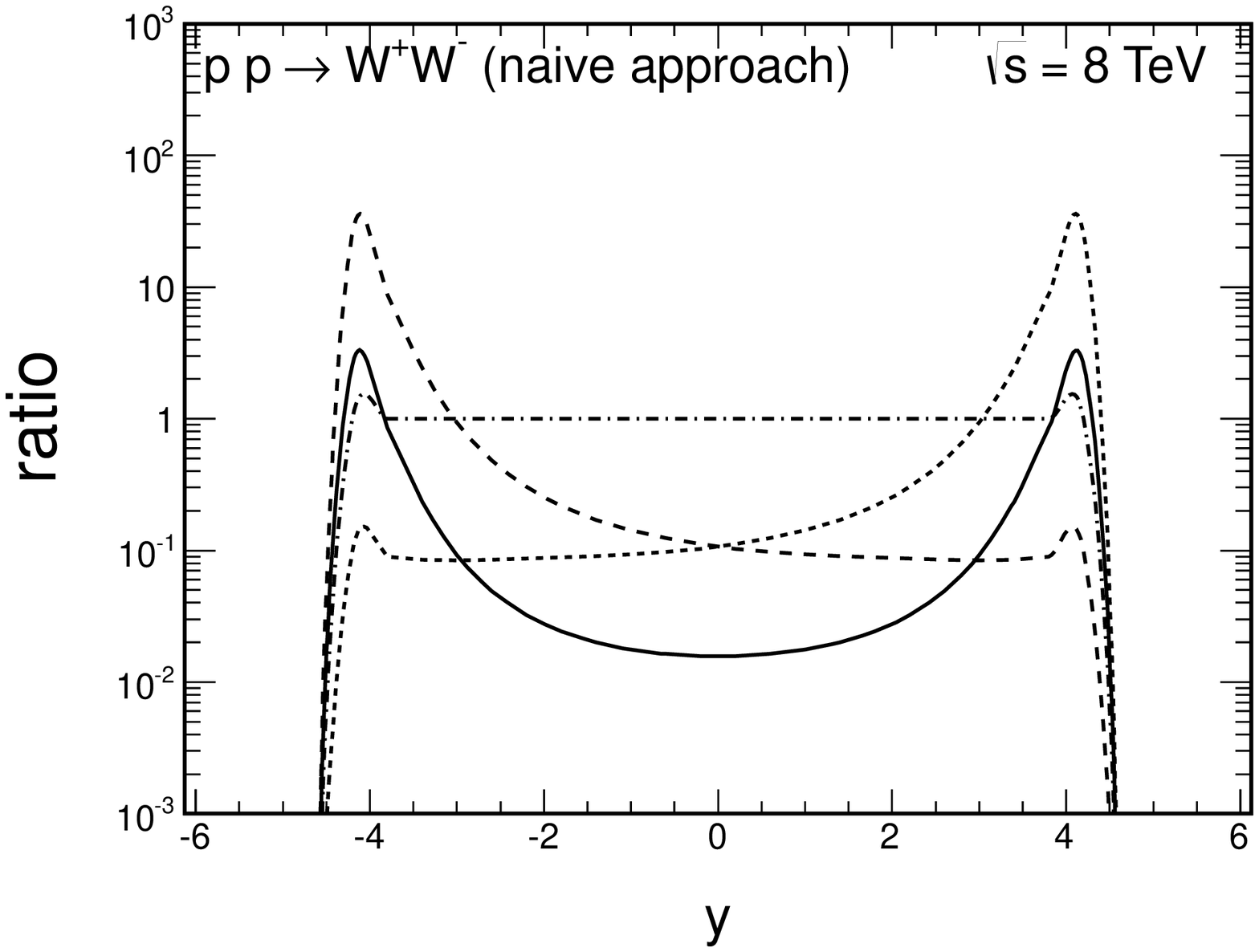}
\includegraphics[width=8cm,height=5cm]{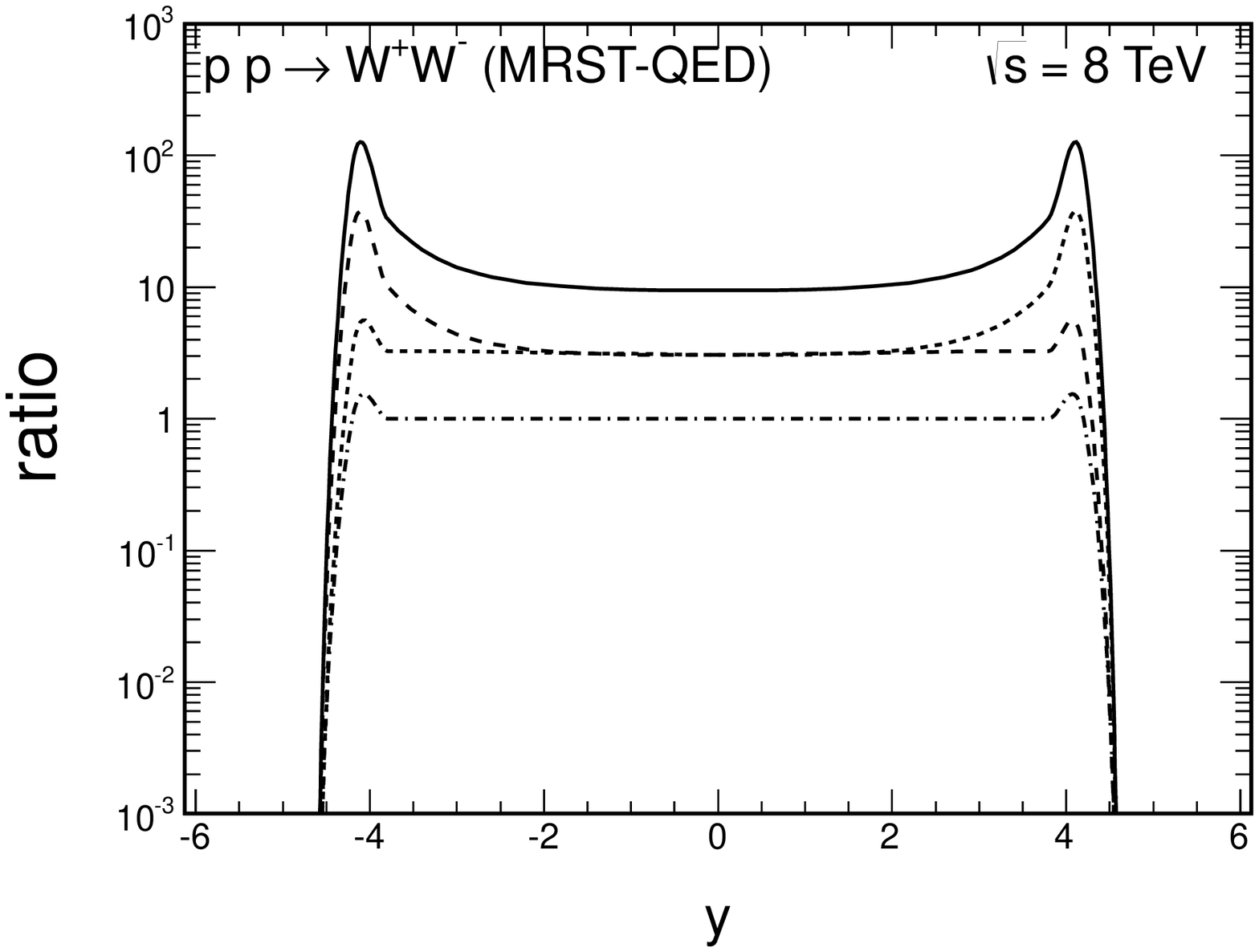}
\end{center}
\caption {Ratio of elastic-inelastic (dashed), inelastic-elastic
  (dotted), inelastic-inelastic (dash-dotted) and all inelastic (solid)
to the elastic-elastic cross section as a function of the $W$ boson
rapidity in the naive (left panel) and QCD improved (right panel) approaches.
}
\label{fig:ratio_y}
\end{figure}

\begin{figure}
\begin{center}
\includegraphics[width=8cm,height=5cm]{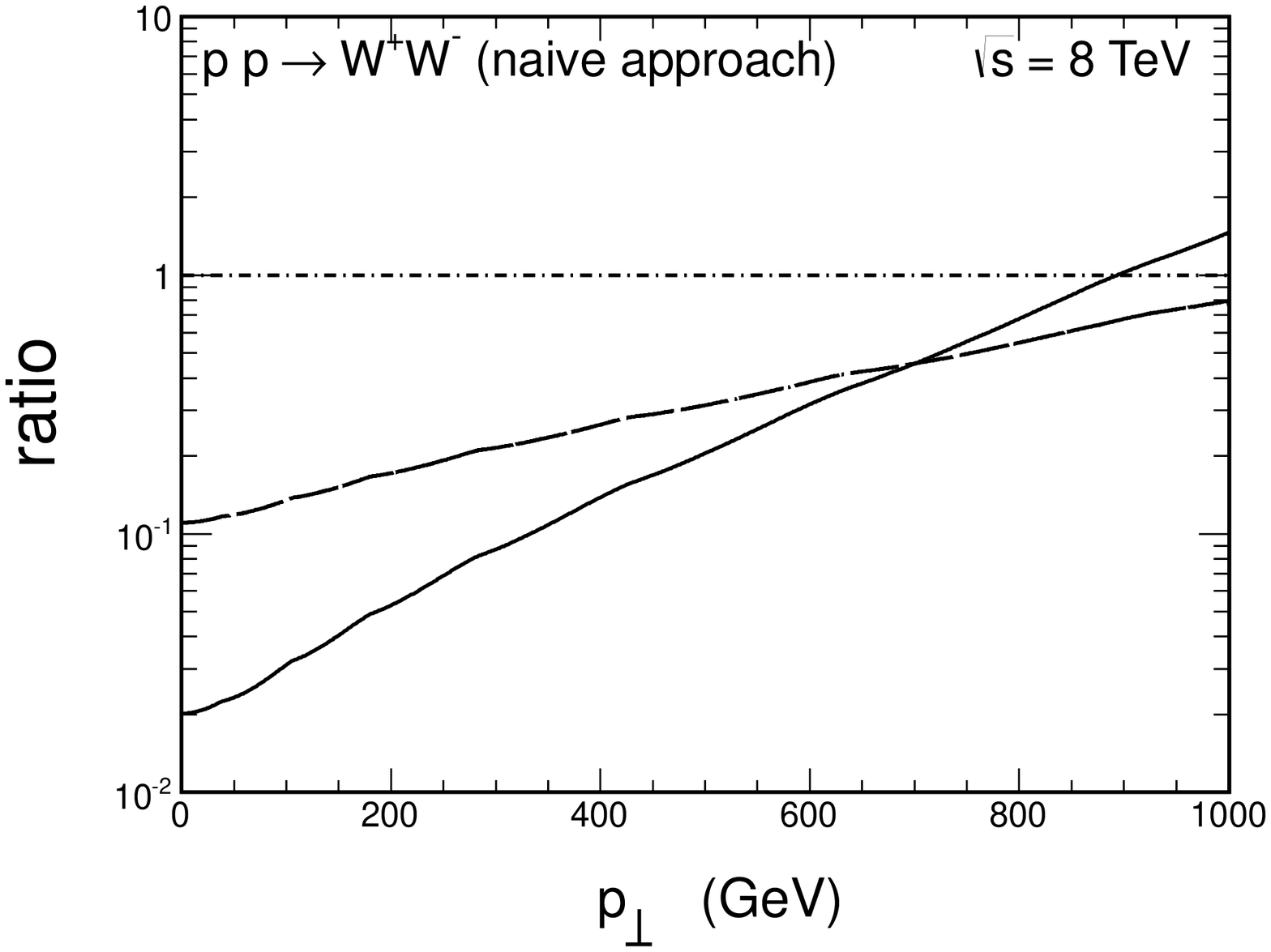}
\includegraphics[width=8cm,height=5cm]{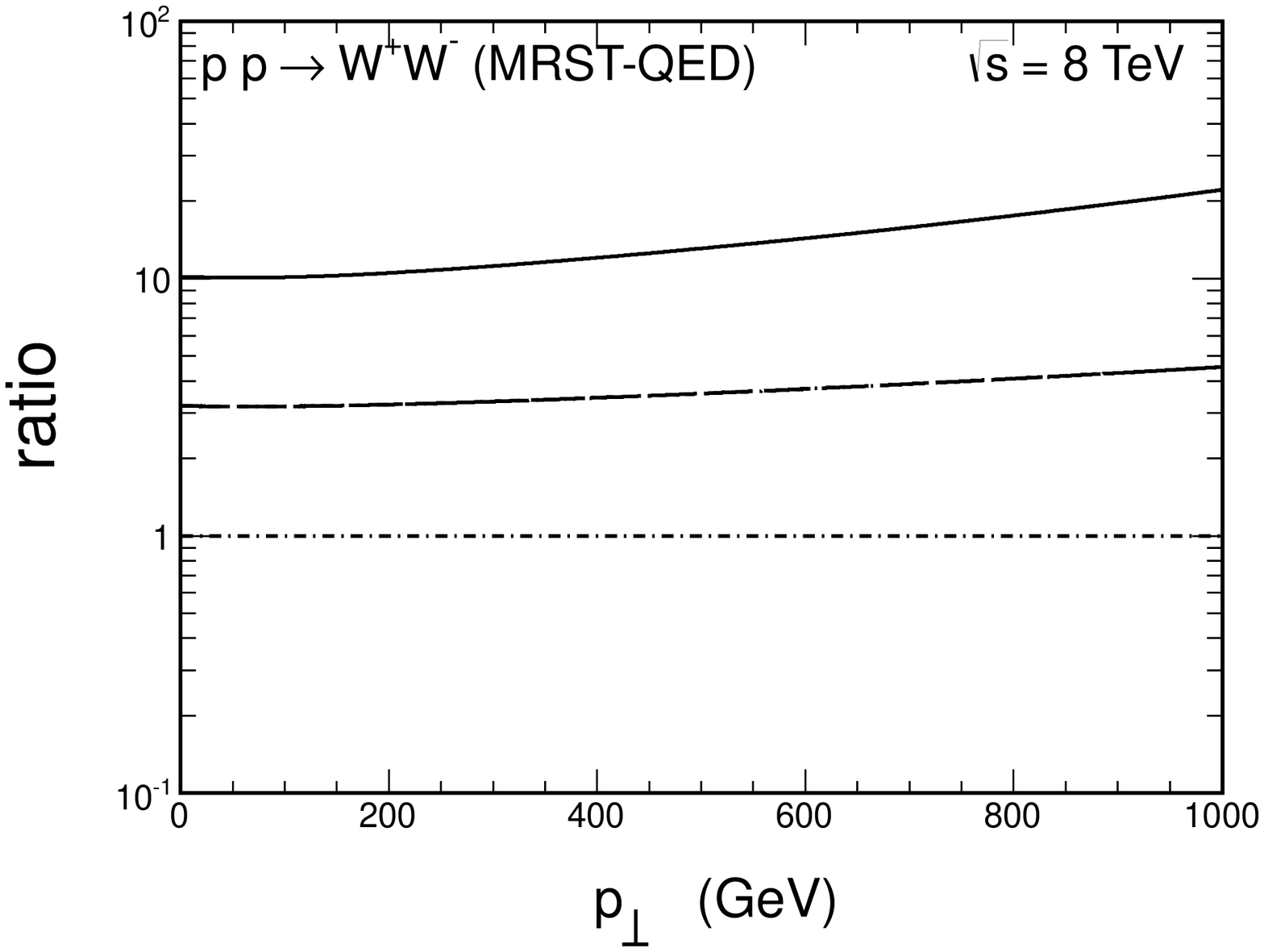}
\end{center}
\caption {Ratio of elastic-inelastic (dashed), inelastic-elastic
  (dotted), inelastic-inelastic (dash-dotted) and all inelastic (solid)
to the elastic-elastic cross section as a function of the $W$ boson
transverse momentum in the naive (left panel) and QCD improved (right panel) approaches.
}
\label{fig:ratio_pt}
\end{figure}

Concerning the searches of anomalous $\gamma \gamma W W$ coupling
without proton tagging as performed by the D0 and CMS collabrations, the ratios
of the inelastic-inelastic, elastic-inelastic and inelastic-elastic contribution
to the elastic-elastic one are fundamental.
In Fig.\ref{fig:ratio_y} and \ref{fig:ratio_pt} we show such ratios in
$W$ boson rapidity and transverse momentum. In the naive approach
the ratios are smaller than 1 except for some small corners of the phase-space.
In the QCD improved approach the ratios become much larger.
In particular, the ratio for inelastic-inelastic contribution is one order of
magnitude larger than 1, showing the importance of tagging the intact protons in
the final state in order to measure the $WW$ exclusive cross section.

\subsection{Discussion of the contributions leading 
to $WW+X$ in the final state}

In this section, we present a systematic survey of all the contributions 
discussed in the present paper. 

\begin{figure}
\begin{center}
\includegraphics[width=6cm,height=5cm]{fig_10b.eps}\\
\includegraphics[width=6cm,height=5cm]{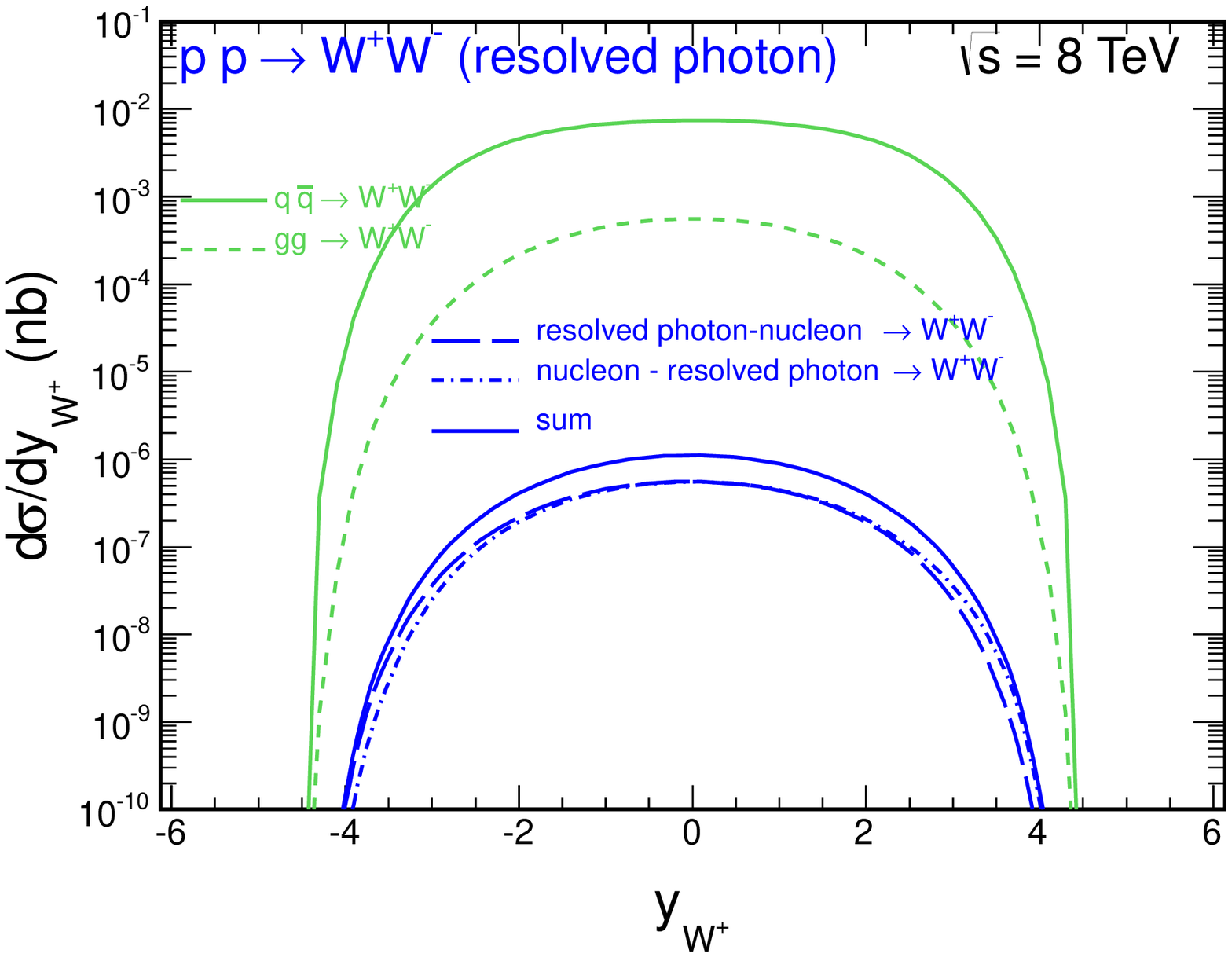}
\includegraphics[width=6cm,height=5cm]{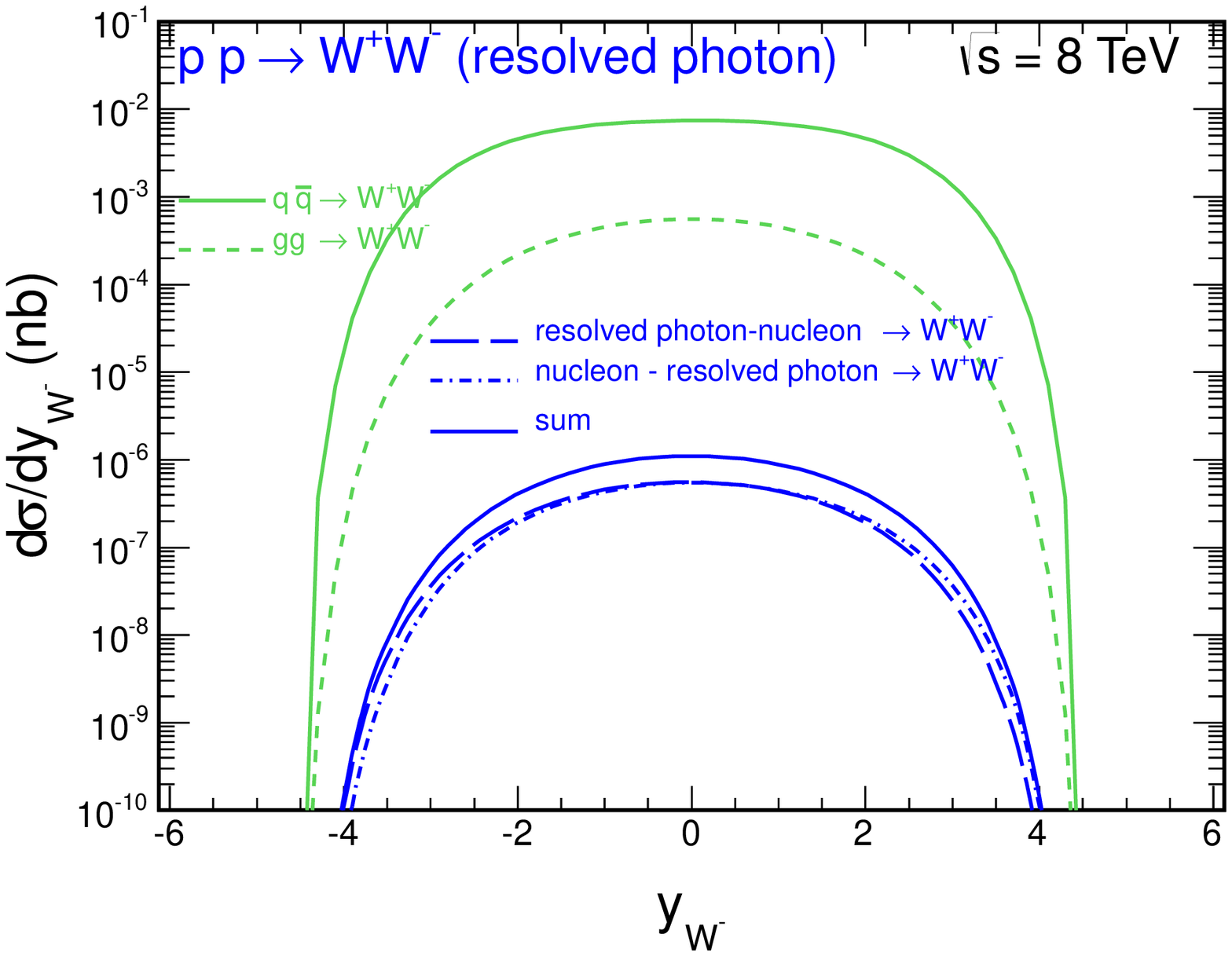}\\
\includegraphics[width=6cm,height=5cm]{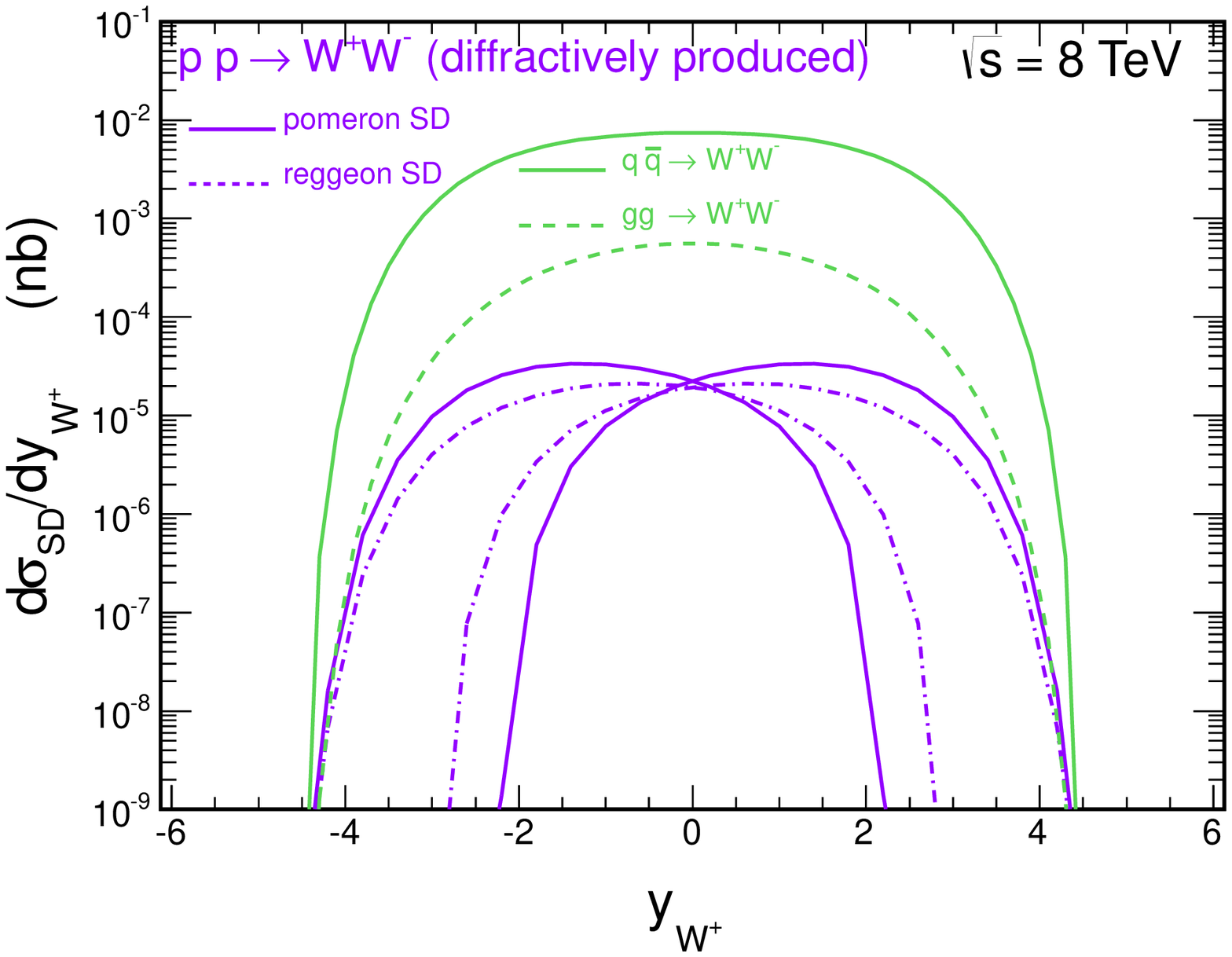}
\includegraphics[width=6cm,height=5cm]{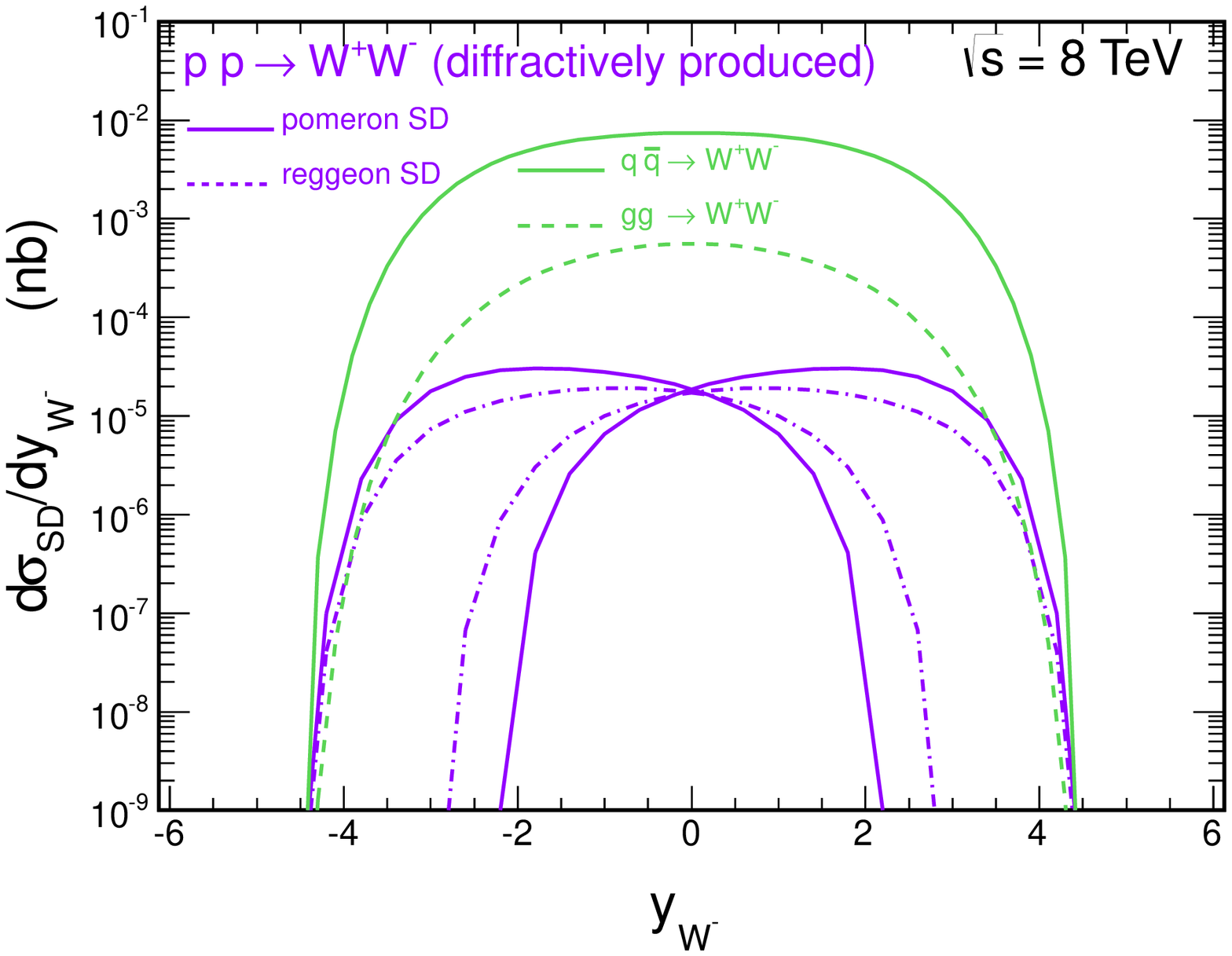}
\end{center}
\caption {$W$ boson rapidity distribution for $\sqrt{s}$ = 8 TeV.
The top panel shows contributions of all photon-photon induced processes
as a function of $y$, the
middle panels resolved photon contributions and the bottom panels 
distributions of the diffractive contribution. The diffractive
cross section has been multiplied by the gap survival factor
$S_G^2$ = 0.08 as needed for requirement of rapidity gaps.
}
\label{fig:dsig_dy}
\end{figure}

The $W$ boson rapidity distribution is shown in
Fig.\ref{fig:dsig_dy}.
We show the separate contributions discussed in the present paper.
The diffractive contribution is an order of magnitude larger than
the resolved photon contribution. The estimated reggeon contribution is
of similar size as the pomeron contribution.
The $y$ distributions of $W^+$ and $W^-$ for double-parton scattering 
contribution are different and, in 
the approximation discussed here, 
have identical shapes as for single production of $W^+$ and $W^-$, 
respectively.
It would be therefore interesting to obtain experimentally separate distributions
for $W^+$ and $W^-$. This is, however, a rather difficult task.
The distributions of charged leptons (electrons, muons) could also be
interesting in this context. It is also worth noticing that the relative size of
each contribution will strongly depend on the mass of the $W$ pair.

\begin{figure}
\begin{center}
\includegraphics[width=8cm,height=5cm]{fig_11b.eps}
\includegraphics[width=8cm,height=5cm]{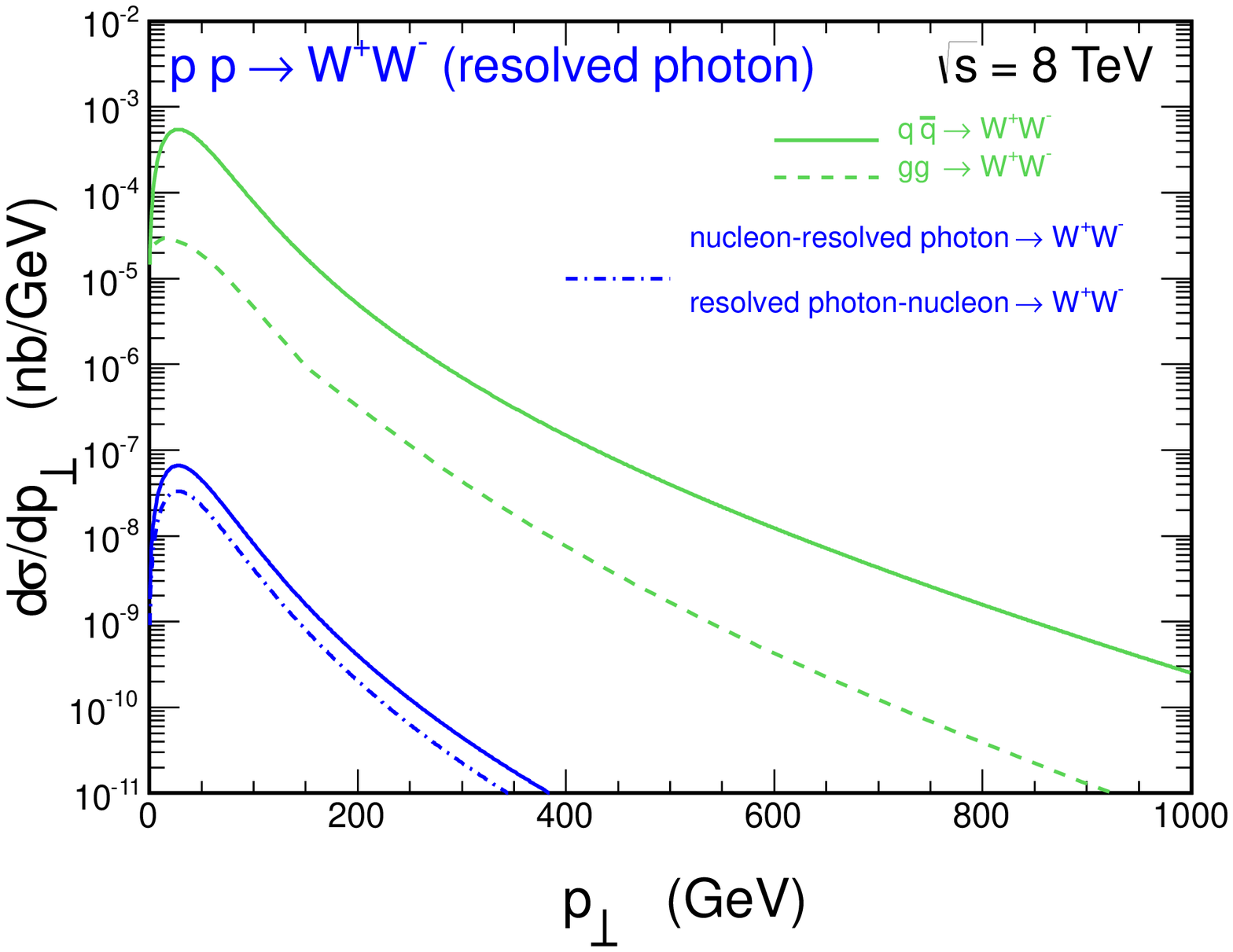}
\includegraphics[width=8cm,height=5cm]{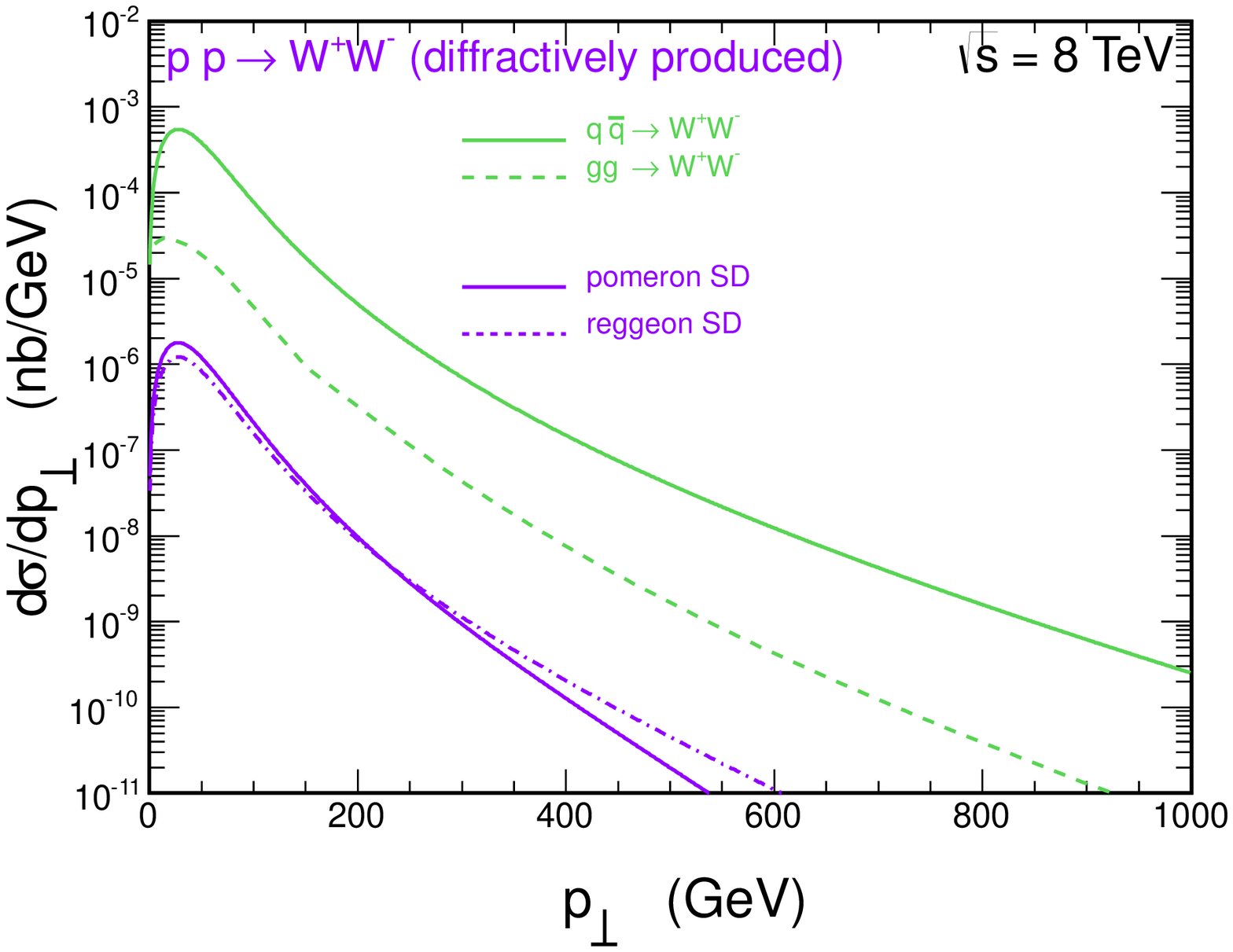}
\end{center}
\caption{$W$ boson transverse momentum distribution for 
$\sqrt{s}$ = 8 TeV.
The left-top panel shows all photon-photon induced processes, the
right-top panel resolved photon contributions and the bottom panel the
diffractive contribution.
The diffractive cross section has been multiplied by the gap survival
factor $S_G^2 =$ 0.08.
The distributions are of similar shape
except for the $\gamma \gamma$ one which is slightly harder.
}
\label{fig:dsig_dpt}
\end{figure}

In Fig.\ref{fig:dsig_dpt} we present the distributions in transverse momentum
of the $W$ bosons. All photon-photon components have rather similar shapes.
The photon-photon contributions are somewhat harder (less steep) than those for
diffractive and resolved photon mechanisms. 

%

\begin{figure}
\begin{center}
\includegraphics[width=6cm,height=5cm]{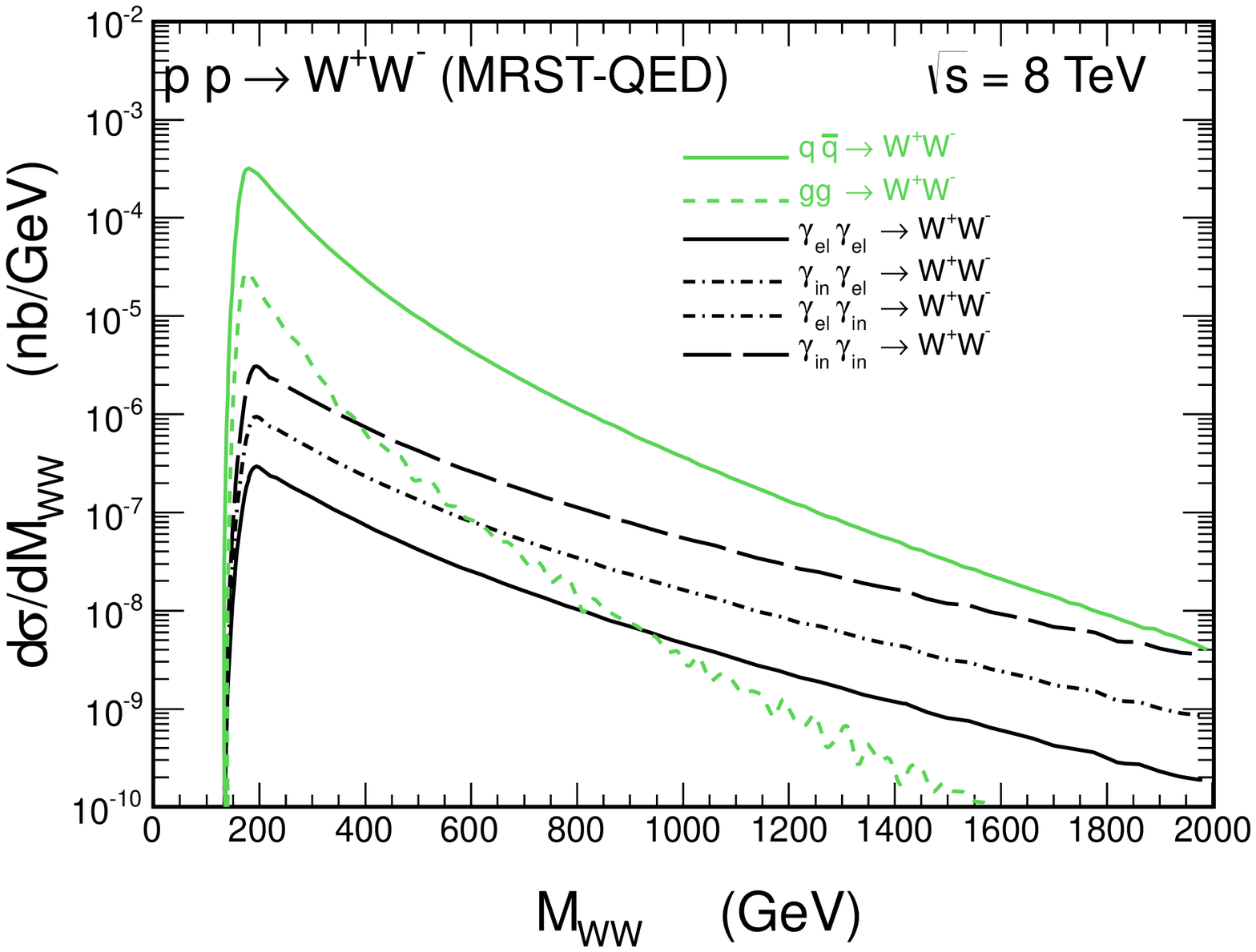}
\includegraphics[width=6cm,height=5cm]{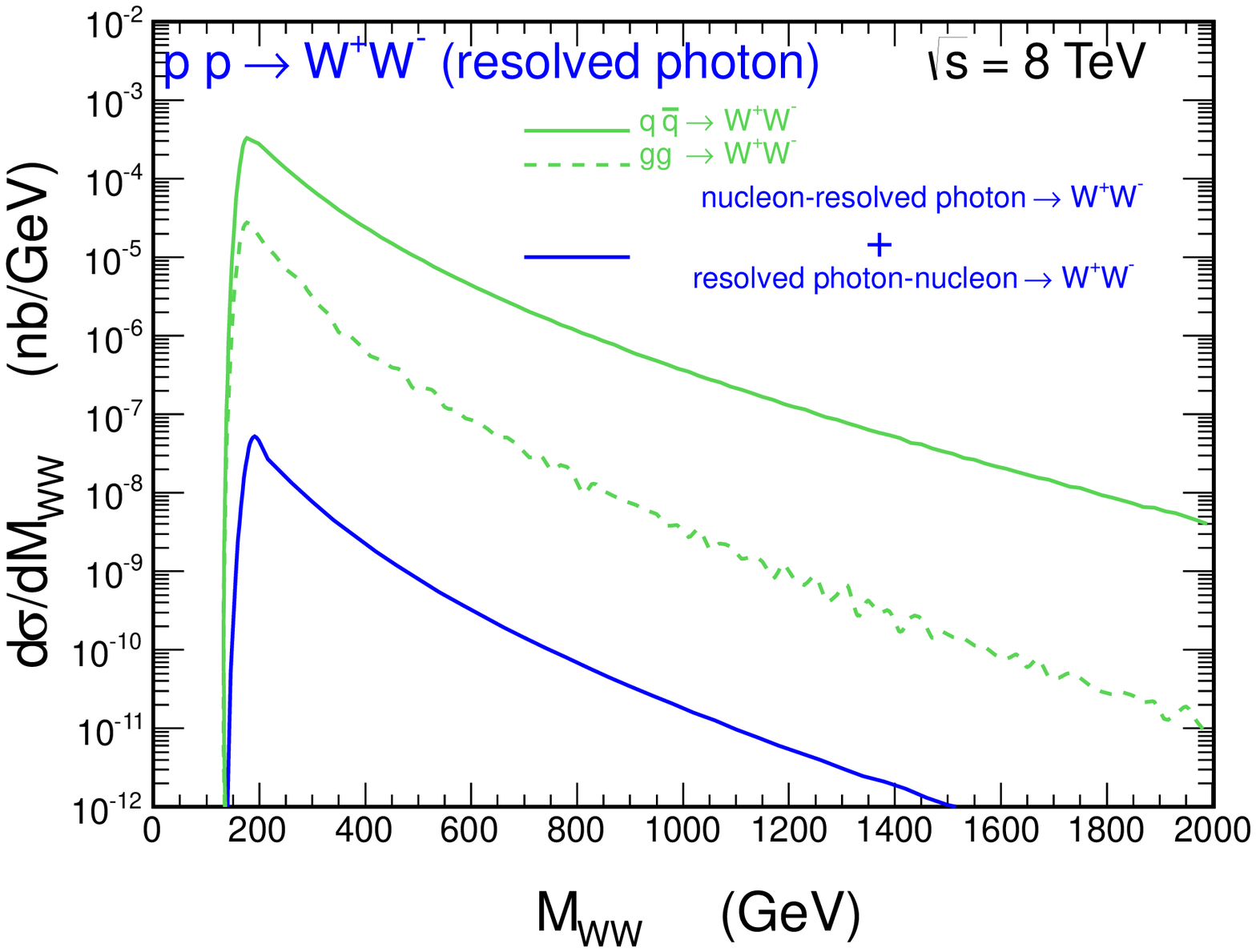}
\includegraphics[width=6cm,height=5cm]{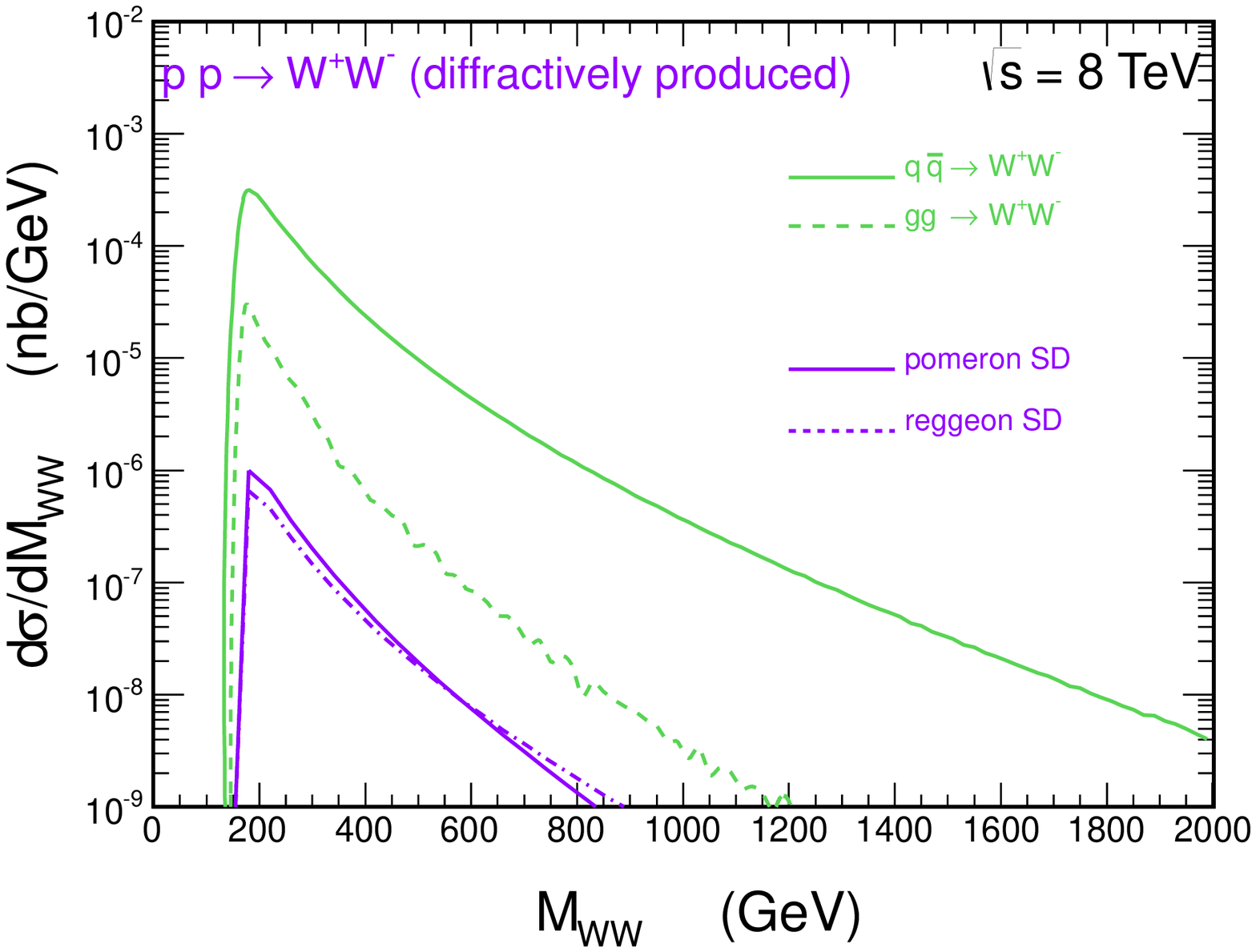}
\includegraphics[width=6cm,height=5cm]{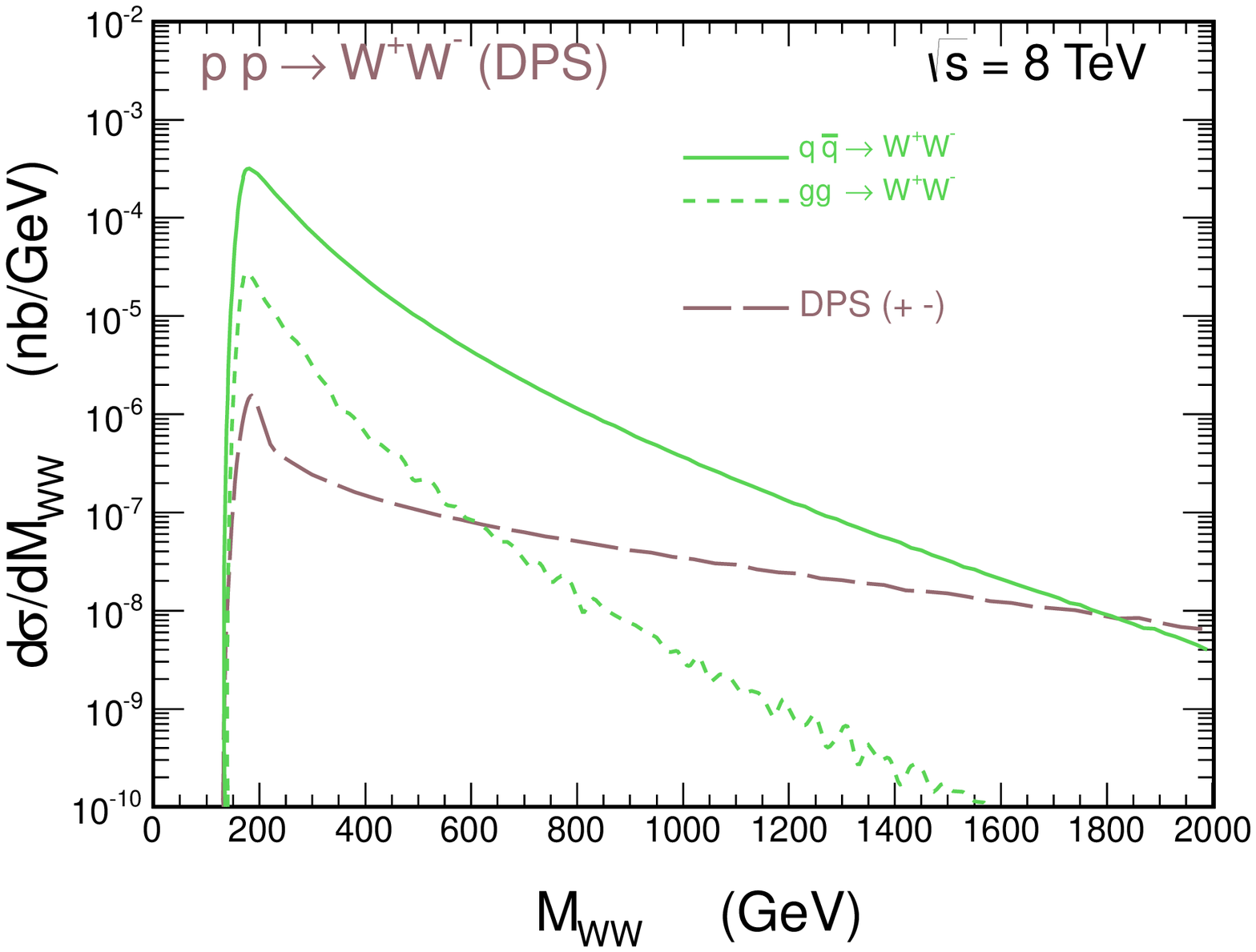}
\end{center}
\caption{$WW$ invariant mass distribution for 
$\sqrt{s}$ = 8 TeV.
The top-left panel shows contributions of all photon-photon induced 
processes, the top-right panel resolved photon contributions, 
the bottom-left panel the contributions of diffractive processes and 
the bottom-right panel the DPS contribution.
The diffractive cross section has been multiplied by the gap survival
factor $S_G^2$ = 0.08.
The $\gamma \gamma$ contribution is larger
at large masses than the DPS one.
}
\label{fig:dsig_dM}
\end{figure}

In Fig.\ref{fig:dsig_dM} we show the distributions in invariant mass of the
$W $ pairs. The relative contributions of the photon-photon
and DPS components grow with the invariant mass. 
The same conclusion is true for the distribution in the rapidity distance 
between the gauge bosons.
Experimentally, one rather measures the charged leptons than W bosons. Measuring the distributions
in invariant mass of positive and negative leptons or in rapidity
distance between them would be interesting.

\begin{figure}
\begin{center}
\includegraphics[width=5cm,height=5cm]{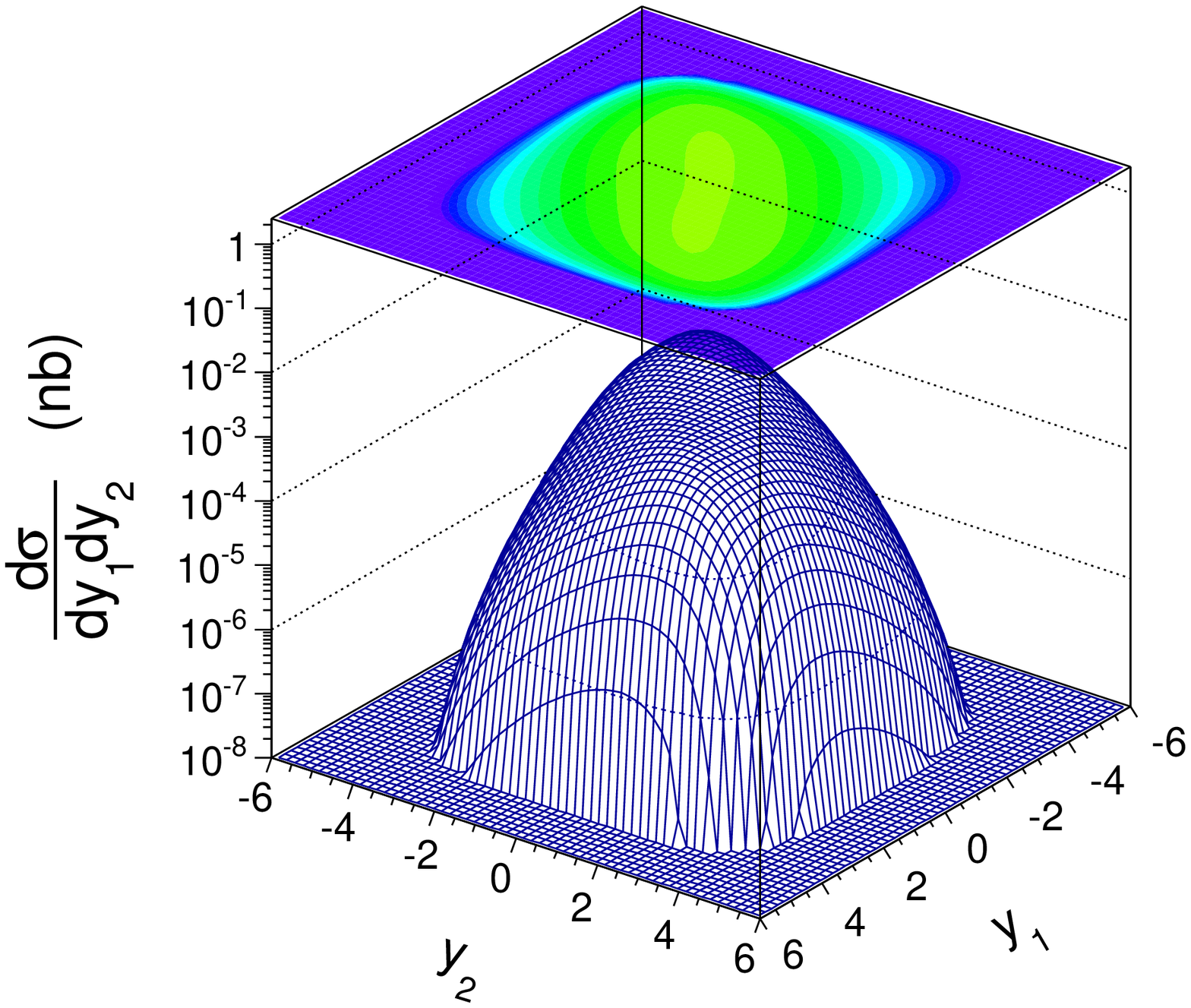}
\includegraphics[width=5cm,height=5cm]{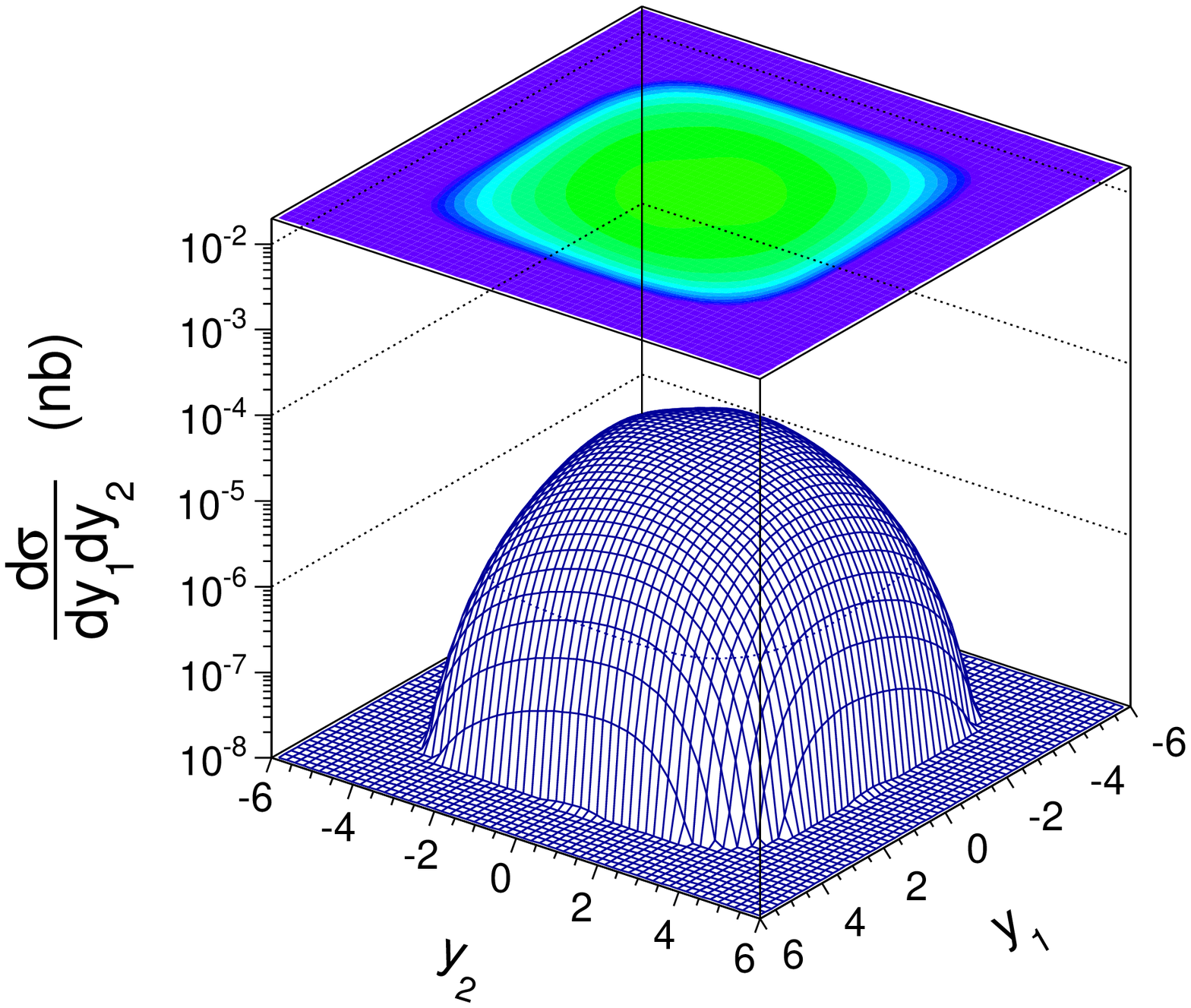}
\includegraphics[width=5cm,height=5cm]{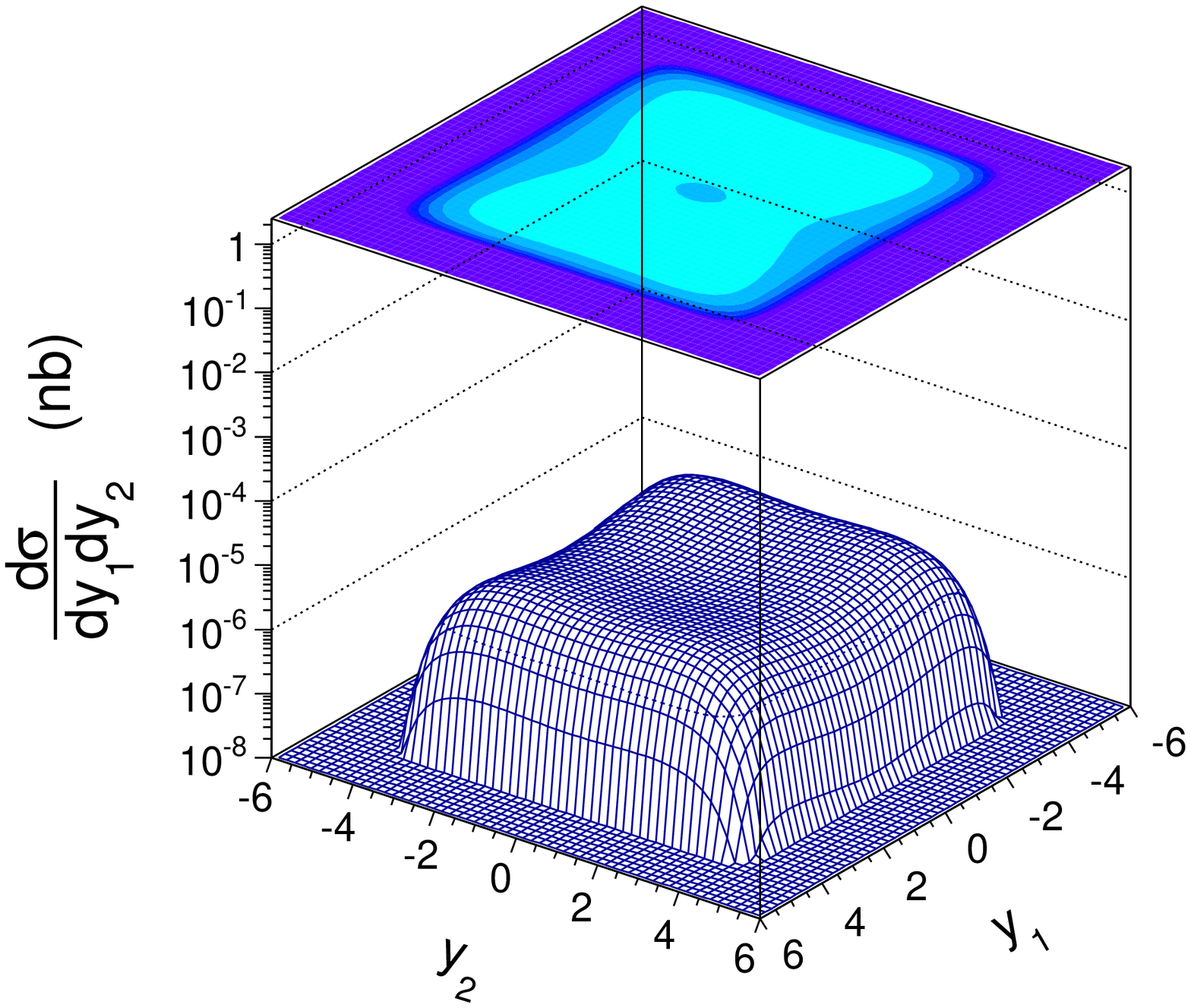}
\end{center}
\caption{ 
Two-dimensional distributions in rapidity of $W^+$ and rapidity of $W^-$
for the dominant $q \bar q$ (left), inelastic-inelastic photon-photon
(middle) and double-parton scattering (right) contributions at 
$\sqrt{s}$ = 8 TeV.}
\label{fig:map_y1y2}
\end{figure}

Finally, in Fig.\ref{fig:map_y1y2} we present some 
interesting examples of two-dimensional distributions in  
$W^+$ and $W^-$ rapidity. We show three different distributions for the dominant
$q \bar q$, inelastic-inelastic photon-photon and double-parton
scattering components. The $q \bar q$ component dominates
at $y_1, y_2 \approx$ 0. The photon-photon component has ``broader''
distribution in $y_1$ and $y_2$. In contrast, the double-parton
scattering component gives a very flat two-dimensional distribution.
The information presented in the figure can be used in order
to ``enhance'' each component. The study of the lepton rapidity
from the $W$-boson decays would be a next interesting step.

The results are summarized in Table 1. The photon-photon
induced processes give quite large contribution. The
single-resolved photon contributions are at least one order of
magnitude smaller than the diffractive one.
This is still surprisingly large. The reason that the single resolved
photon contributions are relatively large is due to the fact that
quark or antiquark carry on average a fairly large fraction of the photon
longitudinal momentum. The double parton scattering contribution is
small. It may be, however, important for large rapidity
distances between gauge bosons. 
The diffractive contributions in the table are
multiplied by the gap survival factor ($S^2_G$) which is known 
only approximately. The cross section of the additional processes usually not
considered in the litterature are thus not enough to explain the discrepancy
between data and theoretical calculations. New data at 14 TeV will be of
considerable interest to know if this discrepancy is real.

In real experiments only a rather limited part of the phase space is covered.
The total cross section is then obtained by extrapolating into unmeasured
region with the help of Monte Carlo codes.
It is needless to say that none of the new contributions discussed here 
is included when extrapolating purely experimental results to 
the phase-space integrated cross sections. 
This means that the ``measured'' total cross section is underestimated.
Answering the question ``by how much'' requires dedicated 
Monte Carlo analyses.

\begin{table}[tbp]
\centering
\begin{tabular}{|c|c|c|c|c|c|}
\hline
contribution               &  1.96 TeV & 7 TeV & 8 TeV & 14 TeV &  comment \\
\hline
CDF                        &  12.1 pb  &       &       &        & \\
D0                         &  13.8 pb &       &       &        & \\
ATLAS                      &       &  54.4 pb  &    &        &  large extrapolation \\
CMS                        &       &  41.1 pb  &    &        &  large extrapolation \\
\hline
$q \bar q$  & 9.86  &  27.24  &   33.04  & 70.21  & dominant (LO, NLO) \\
$g g$     & 5.17 10$^{-2}$  & 1.48  &  1.97  & 5.87 &  subdominant (NLO) \\
\hline        
$\gamma_{el} \gamma_{el}$  & 3.07 10$^{-3}$ & 4.41 10$^{-2}$ & 5.40 10$^{-2}$ & 1.16 10$^{-1}$ &  new, anomalous $\gamma \gamma W W$  \\
$\gamma_{el} \gamma_{in}$  & 1.08 10$^{-2}$ & 1.40 10$^{-1}$ & 1.71 10$^{-1}$ & 3.71 10$^{-1}$ &  new, anomalous $\gamma \gamma W W$  \\
$\gamma_{in} \gamma_{el}$  & 1.08 10$^{-2}$ & 1.40 10$^{-1}$ & 1.71 10$^{-1}$ & 3.71 10$^{-1}$ &  new, anomalous $\gamma \gamma W W$  \\
$\gamma_{in} \gamma_{in}$  & 3.72 10$^{-2}$ & 4.46 10$^{-1}$ & 5.47 10$^{-1}$ & 1.19  &  anomalous $\gamma \gamma W W$ \\
\hline
$\gamma_{el,res} - q/\bar q$ & 1.04 10$^{-4}$  & 2.94 10$^{-3}$  & 3.83 10$^{-3}$ & 1.03 10$^{-2}$ &  new, rather sizeable \\ 
$q/\bar q - \gamma_{el.res}$ & 1.04 10$^{-4}$  & 2.94 10$^{-3}$  & 3.83 10$^{-3}$ & 1.03 10$^{-2}$ &  new, rather sizeable \\    
$\gamma_{in,res} - q/\bar q$ &       &       &       &        &  not calculated \\ 
$q/\bar q - \gamma_{in.res}$ &       &       &       &        &  not calculated \\    
\hline
DPS(++)     &  0.61 10$^{-2}/2$     & 0.22/2              & 0.29/2     &    1.02/2  &  not included in NLO \\
DPS(- -)    &  0.58 10$^{-2}/2$     & 0.76 10$^{-1}/2$    & 0.11/2     &    0.40/2  &  not included in NLO \\
DPS(+-)     &  0.6 10$^{-2}$     & 0.13              & 0.18     &    0.64  &  not included in NLO \\
DPS(-+)     &  0.6 10$^{-2}$     & 0.13              & 0.18     &    0.64  &  not included in NLO \\
\hline
$\Pom p$  $(x_{\Pom} < 0.1)$     &  0.28 10$^{-2}$    &  0.79 10$^{-1}$   &  0.11    &   0.27      &  new, relatively large \\
$p \Pom$  $(x_{\Pom} < 0.1)$     &  0.28 10$^{-2}$    &  0.79 10$^{-1}$   &  0.11    &   0.27      &  new, relatively large \\
$\Reg p$  $(x_{\Reg} < 0.2)$     &  0.45 10$^{-2}$    &  0.57 10$^{-1}$   & 0.72 10$^{-1}$  &  0.18   &  new, relatively large \\
$p \Reg$  $(x_{\Reg} < 0.2)$     &  0.45 10$^{-2}$    &  0.57 10$^{-1}$   & 0.72 10$^{-1}$  &  0.18   &  new, relatively large \\
\hline
all subleading contributions     &  8.87 10$^{-2}$  &  1.31   &  1.68   &  4.25      &   \\
\hline
\end{tabular}
\caption{Contributions of different subleading processes
  discussed in the present paper to the $WW+X$ cross section
for different energies. The cross section is given in pb. The 
diffractive contributions have been multiplied by the gap
survival factors: 
0.1 at 1.96 TeV, 0.08 at 7 and 8 TeV (see \cite{CMS_2013}) and 0.03 at 14 TeV.
The combinatorial factor 1/2 is shown explicitly for
the double parton scattering (DPS) contribution for $W^+ W^+$ and 
$W^- W^-$ final states shown for completeness.}
\end{table}

Some comments on recent studies on $\gamma \gamma W^+ W^-$ boson
couplings, as performed recently by the D0 and CMS collaborations
\cite{gamgam_WW_D0,gamgam_WW_CMS} are in order. In the D0 collaboration 
analysis the inelastic contributions are not included when extracting 
limits on anomalous couplings, that lead to conservative limits.
The CMS collaboration requires an extra condition of no charged
particles in the central pseudorapidity interval.
When comparing calculations to the experimental data the inelastic
contributions are estimated by rescaling the elastic-e
lastic contribution
by an experimental function depending on kinematical variables
(invariant mass, transverse momentum of the $\mu^+ \mu^-$ pair)
obtained in the analysis of the $\mu^+ \mu^-$ continuum.
It is not clear to us whether such a procedure is consistent for
$W^+ W^-$ production, where leptons come from the decays of the gauge
bosons and the invariant mass and transverse momentum of the $W^+ W^-$ pair
is very different than the invariant mass and transverse momentum
of the corresponding dimuons. This cannot be checked in the present 
approach with collinear photons and requires the inclusion of photon
transverse momenta.
Within the present approach we predict that the inelastic $\gamma \gamma$ contributions
are significantly larger than the elastic-elastic one. The fragmentation
of the remnants of the inelastic excitations is needed to understand to
which extent the inelastic contributions survive the veto condition.
One can expect that particles from the fragmentation of the proton remnants
after photon emission are emitted in rather forward/backward directions.
The $D0$ collaboration also neglects the inelastic contribution when
calculating the Standard Model background at large lepton transverse 
momenta, despite they have no explicit veto condition on charged
particles. 
The diffractive contributions could also contribute to the distributions 
measured by the CMS and D0 collaborations.
Clearly further analyses that focus on final states of proton remnants 
(after photon emission) are necessary.

\section{Conclusions}

In the present paper we discussed several contributions not
considered traditionally
in the so-called Standard Model predictions such as:
photon-photon induced processes, processes with resolved photons, 
single and central diffractive processes and double parton scattering.

We have calculated for the first time the complete set of
photon-photon  and resolved photon-(anti)quark and (anti)quark-resolved 
photon contributions to the inclusive production of $W^+ W^-$ pairs. 
The photon-photon contributions can be classified into four topological
categories: elastic-elastic, elastic-inelastic, inelastic-elastic
and inelastic-inelastic, depending whether the proton(s) survives 
the emission of the photon or not. The elastic-inelastic and
inelastic-elastic contributions were calculated for the first time in this study.
The photon-photon contributions were calculated as in previous studies. 
for production of pairs of charged Higgs bosons or pairs of heavy
leptons beyond the Standard Model, and within QCD-improved method 
using the MRST(QED) parton distributions. 
The second approach was already applied to the production of
Standard Model charged lepton pair production
and $c \bar c$ production.
In the more refined approach we got
$\sigma_{ela,ela} <\sigma_{ela,ine} = \sigma_{ine,ela} <
\sigma_{ine,ine}$.
In the approach when photon distribution in the proton undergoes QCD 
$\otimes$ QED evolution, the inelastic-inelastic contribution 
is the largest out of the four contributions.
This shows that including photon into evolution equation is crucial.
This includes also some processes beyond the Standard Model mentioned in
this paper.

The inelastic contributions sum up to the cross section of the order 
of 0.5 - 1 pb at the LHC energies. The photon-photon contributions
are particularly important at large $WW$ invariant masses, i.e. 
probably also large invariant masses of charged leptons where 
its contribution is larger than that for gluon-gluon fusion.

The elastic-inelastic or inelastic-elastic contributions are 
interesting by themselves. Since they are related to the emission of 
forward/backward protons they could be potentially measured in the future
with the help of forward proton detectors.
Both CMS and ATLAS have plans for installing such detectors after
the present (2013-2014) shutdown.
Unfortunately the elastic-inelastic mechanisms are expected to have similar topology of 
the final state as single-diffractive contributions to $W^+ W^-$ production. 
It would be therefore valuable to make a dedicated stu
dy how to pin down 
the mixed elastic-inelastic contributions. Clearly this would be a
valuable test of both the presented formalism and our understanding 
of the underlying reaction mechanism.

We have also discussed briefly the double-scattering mechanism
which also significantly contributes to large $M_{WW}$ invariant masses,
the region where beyond Standard Model processes can be expected.
The DPS was suggested recently as an important ingredient
for the Higgs background in the $W W^*$ or $Z Z^*$ final channels.
Our estimate is more than order of magnitude smaller than the one
suggested recently in the litterature in order to explain the Higgs
signal in the $W^+ W^-$ channel. However, the DPS calculation
contains some phenomenological ingredients which are not fully
understood.

Summarizing, we have shown that the omitted so far processes
can contribute a few pb at $\sqrt{s}$ = 8 TeV to the
inclusive $W^+ W^-$ production cross section. There are 
even larger than at 14 TeV.
The contribution of the subleading processes is growing
faster with increasing center of mass energy than the standard 
Standard Model contributions (see the last row in the 
table).
As discussed in our paper a precise estimation of contribution 
of these, also Standard Model, processes is rather difficult and
requires further work.
Some of the contributions discussed here have specific final
state topology with one forward/backward proton separated in rapidity.
Measuring the final state would therefore help to identify elastic-inelastic
(inelastic elastic) two-photon contributions.
Also single-diffractive contribution should have similar topology.
They should have, however different dependence on t 
(four-momentum transfer squared).

After this paper was completed we have learned about a detailed study
of uncertainties of a photon PDF in the framework of so-called neural
network PDFs \cite{NNPDF}. This analysis suggests that inelastic
photon-induced contributions may have rather big uncertainies. 
The issue should be better clarified in the future by comparing 
similar calculation for two-photon-induced $\mu^+ \mu^-$ production
at large invariant masses of dileptons to appropriate experimental data.

\vspace{1cm}

{\bf Acknowledgments}

We are indebted to Krzysztof Piotrzkowski, Jonathan Hollar, 
and Gustavo da Silveira for a discussion of the CMS experiment 
on semi-exclusive production of $W^+ W^-$ pairs. 
The help of Piotr Lebiedowicz in calculating the gluon-gluon component
is acknowleded. We are indebted to Wolfgang Sch\"afer for a FORTRAN routine
and an interesting discussion. We are indebted to Stefano Forte and
Juan Rojo for a communication on their recent work on photon NNPDF.
This work was partially supported by the Polish NCN grants: 
DEC-2011/01/B/ST2/04535 and DEC-2013/09/D/ST2/03724 
and by the bilateral exchange program
between Polish Academy of Sciences and FNRS, Belgium.



\end{document}